\documentclass[11pt,a4paper]{article}
\usepackage{jheppub}
\pdfoutput=1

\usepackage[T1]{fontenc}
\usepackage{amssymb}
\usepackage{amsmath}
\usepackage{latexsym}
\usepackage{color}
\usepackage[mathscr]{eucal}
\usepackage{fancybox}
\usepackage{simplewick}
\usepackage{comment}
\usepackage{bm}
\usepackage{framed}
\definecolor{shadecolor}{rgb}{0.9,0.9,0.95}
\usepackage{setspace}
\usepackage[normalem]{ulem}
\usepackage{orcidlink}
\definecolor{darkgreen}{rgb}{0,0.5,0}
\definecolor{darkblue}{cmyk}{0.9,0.9,0,0}
\definecolor{darkred}{rgb}{0.6,0,0.3}

\usepackage{graphicx}
\usepackage[curve,matrix,arrow,color]{xy}
\usepackage{tikz}
\usetikzlibrary{intersections}
\usetikzlibrary{er,positioning}
\usetikzlibrary{arrows,shapes}
\usetikzlibrary{decorations.markings}
\usepackage{hyperref}

\renewcommand{\thefootnote}{\arabic{footnote}}

\def\eqref#1{(\ref{#1})}

\def\beq{\begin{equation}}
\def\eeq{\end{equation}}

\renewcommand{\thefigure}{\thesection.\arabic{figure}}

\newcommand{\rd}{\mathrm{d}}

\numberwithin{equation}{section}

\begin{document}

\begin{flushright}
{\tt USTC-ICTS/PCFT-23-35
}
\end{flushright}
\title{Giant Correlators at Quantum Level}

\author[a]{Yunfeng Jiang}
\author[b]{Yu Wu}
\author[b, c]{Yang Zhang\orcidlink{0000-0001-9151-8486}}
\affiliation[a]{School of physics \& Shing-Tung Yau Center, Southeast University,\\ Nanjing  211189, P. R. China}
\affiliation[b]{Interdisciplinary Center for Theoretical Study, University of Science and Technology of China,\\ Hefei, Anhui 230026, China}
\affiliation[c]{Peng Huanwu Center for Fundamental Theory, Hefei, Anhui 230026, China}

\abstract{
We compute four-point functions with two maximal giant gravitons and two chiral primary operators at three-loop order in planar $\mathcal{N}=4$ Super-Yang-Mills theory. The Lagrangian insertion method, together with symmetries of the theory fix the integrand up to a few constants, which can be determined by lower loop data. The final result can be written in terms of the known three-loop conformal integrals. From the four-point function, we extract the OPE coefficients of two giant gravitons and one non-BPS twist-2 operator with arbitrary spin at three-loops, given in terms of harmonic sums. We observe an intriguingly simple relation between the giant graviton OPE coefficients and the OPE coefficients of three single-trace operators.}
\emailAdd{101013107@seu.edu.cn}
\emailAdd{wy626@mail.ustc.edu.cn}
\emailAdd{yzhphy@ustc.edu.cn}
\maketitle

\renewcommand{\thefootnote}{\fnsymbol{footnote}}
\setcounter{page}{1}
\setcounter{footnote}{0}
\setcounter{figure}{0}

\setcounter{page}{1}
\renewcommand{\thefootnote}{\arabic{footnote}}
\setcounter{footnote}{0}
\setcounter{tocdepth}{2}
\newpage
\tableofcontents

\section{Introduction}
\label{sec:intro}

Correlation functions play an important role in both quantum field theory and holography. In $\mathcal{N}=4$ Super-Yang-Mills ($\mathcal{N}=4$ SYM) theory, one of the most interesting class of observables are the four-point functions of BPS operators. However, it is well-known that computing them non-perturbatively is extremely challenging. A more pragmatic approach is first computing them perturbatively, both at weak and strong couplings. Such perturbative results are already very useful. They encode rich information of anomalous dimensions and OPE coefficients of local operators. These perturbative data are essential for fixing various ambiguities in the integrability method, which is a non-perturbative approach. At the same time, four-point functions are starting points for the conformal bootstrap program \cite{Poland:2018epd,Poland:2022qrs,Hartman:2022zik}. A better understanding of their analytic properties and perturbative results is important for bootstraping them non-perturbatively \cite{Caron-Huot:2022sdy,Cavaglia:2021bnz,Cavaglia:2022qpg}.\par

A majority of the works on four-point functions of $\mathcal{N}=4$ SYM theory so far have focused on the cases where the four BPS operators are \emph{single-trace operators}. Thanks to the huge symmetry of $\mathcal{N}=4$ SYM, important progress have been made at weak \cite{Eden:2011we,Eden:2012tu,Chicherin:2015edu,Chicherin:2018avq,Bourjaily:2016evz}, strong \cite{DHoker:1999kzh,Arutyunov:2000py,Rastelli:2016nze,Rastelli:2017udc,Alday:2019nin,Alday:2020dtb,Drummond:2019hel,Aprile:2019rep,Drummond:2022dxw,Caron-Huot:2018kta,Caron-Huot:2021usw,Alday:2022uxp,Alday:2022xwz,Alday:2023flc,Alday:2023jdk,Alday:2023mvu,Fardelli:2023fyq} and even finite coupling \cite{Fleury:2016ykk,Fleury:2017eph,Basso:2017khq,Coronado:2018ypq,Coronado:2018cxj} in the planar limit. Non-planar contributions are much more challenging. Nevertheless, $1/N_c$ corrections of four-point functions have also been investigated in \cite{Bargheer:2017nne,Bargheer:2018jvq,Fleury:2019ydf,Bargheer:2019kxb}. In AdS/CFT correspondence, single-trace operators are dual to graviton and the KK modes of the type-II\,B superstring/supergravity theory. However, single-trace operators only constitute part of the theory. There are many other types of objects such as membranes in superstring theory. Such objects are much heavier than the single-trace operators and their study have been crucial in the development of string theory and holography. As local operators, one of the best studied examples of such soliton-like quantities are the giant gravitons which are dual to D-branes \cite{McGreevy:2000cw,Balasubramanian:2001nh,Hashimoto:2000zp}. \par

Studies of correlation functions involving giant gravitons have started in the early days of AdS/CFT \cite{Corley:2001zk} and continued ever since \cite{deMelloKoch:2004crq,Kimura:2007wy,Bissi:2011dc,Caputa:2012yj,Berenstein:2013md,Kimura:2016bzo,deMelloKoch:2019dda,Chen:2019gsb,Chen:2019kgc,Jiang:2019xdz,Jiang:2019zig,Yang:2021hrl,Yang:2021kot,Vescovi:2021fjf}. Most of the studies from CFT side focus on the Born level, where the Wick contractions already exhibit considerable combinatorics complexity. Various methods have been developed \cite{Kimura:2007wy,Corley:2001zk,deMelloKoch:2004crq,Kimura:2016bzo,Berenstein:2013md,Lin:2022wdr,Lin:2022gbu} to handle such difficulties. The merit of these approaches is that some finite $N_c$ effects can be explored. On the other hand, it is quite difficult to take into account quantum corrections and compute these correlation functions at higher loop orders.\par

In recent years, four-point functions involving giant gravitons have been investigated at the quantum level in the planar limit \cite{Jiang:2019xdz,Jiang:2019zig,Vescovi:2021fjf}. One important motivation for computing these correlation functions is that they contain the OPE coefficients of two giant graviton and one non-BPS operator \cite{Jiang:2019xdz,Jiang:2019zig}. These OPE coefficients can be computed by integrability at any coupling using the worldsheet $g$-function approach \cite{Jiang:2019xdz,Jiang:2019zig}. To test integrability predictions, higher loop data from direct field theoretical computations play an important role. At the moment, correlation functions involving two giant gravitons and two \textbf{20}' operators have been computed up to two-loop order \cite{Jiang:2019xdz,Jiang:2019zig}. The two-loop data is already quite useful for testing important quantities such as the boundary dressing phase. For testing other important effects like wrapping corrections, it is necessary to go to higher loop orders. The goal of the current work is to take a solid step forward and compute such correlation functions at three-loop order.\par 

Given the huge progress of correlation functions of single-trace BPS operators, it is a natural question whether similar achievements could be made for correlation functions involving giant gravitons. This question is obviously important given the fundamental role of D-branes in string theory. At the same time, it is also more challenging. In the current work, we use the Lagrangian insertion method and symmetry constraints to fix the form of planar four-point function up to a few coefficients. These coefficients can be fixed by two-loop data, which can be extracted from a lower-loop field theory or integrability calculation. This shows that at least in the planar limit, some techniques for computing the loop level correlation functions of single trace operators can be effectively adapted to the ones involving giant gravitons.\par

The rest of the paper is structured as follows. In section~\ref{sec:LIM}, we present our set-up and review briefly the Lagrangian insertion method. In section~\ref{sec:TQPoly}, we fix the three-loop integrand up to 4 coefficients by using symmetry and planarity. In section~\ref{sec:OPElimit}, we fix the coefficients by considering the OPE limit of the four-point function and using two-loop data. We exact three-loop OPE coefficients of giant gravitons and twist-2 operators in section~\ref{sec:OPE}. We conclude in section~\ref{sec:conclude} and discuss future directions. Conventions and some technical details are given in the appendices.

\section{Set-up and Lagrangian insertion}
\label{sec:LIM}
We are interested in the following four-point function of $\mathcal{N}=4$ SYM theory
\begin{align}
\label{eq:mainCorr}
G_{\{2,2\}}(x_1,\ldots,x_4)=\langle\mathcal{D}(x_1)\mathcal{D}(x_2)\mathcal{O}_2(x_3)\mathcal{O}_2(x_4)\rangle
\end{align}
where $\mathcal{D}(x_i)$ and $\mathcal{O}_2(x_j)$ are the maximal giant graviton and the length-2 chiral primary operator in the \textbf{20}' representation. More explicitly, they are defined as
\begin{align}
\mathcal{D}(x_i)\equiv\det(Y_i^I\phi^I)(x_i),\qquad \mathcal{O}_2(x_j)=Y_j^IY_j^J\text{tr}(\phi^I\phi^J),\qquad i=1,2;\quad j=3,4\,,
\end{align}
where summing over $I,J$ is understood and $Y_k^I$'s are 6-dimensional null vectors $Y_k^IY_k^I=0$. Both $\mathcal{D}(x_i)$ and $\mathcal{O}(x_j)$ are BPS operators with protected scaling dimensions $N$ and $2$.
At weak coupling, $G_{\{2,2\}}$ has the following perturbative expansion
\begin{align}
\label{eq:pertExpand22}
G_{\{2,2\}}={G}_{\{2,2\}}^{(0)}+g^2{G}_{\{2,2\}}^{(1)}+g^4{G}_{\{2,2\}}^{(2)}+g^6{G}_{\{2,2\}}^{(3)}+\cdots\,,
\end{align}
where $g^2$ is the 'tHooft coupling constant defined by $g^2=g_{\text{YM}}^2N_c/(16\pi^2)$. The results $G_{\{2,2\}}^{(0)},G_{\{2,2\}}^{(1)}$ and $G_{\{2,2\}}^{(2)}$ are known, which are summarized in appendix~\ref{app:pert2loop} for the readers' convenience. Our goal is to compute the three-loop result $G_{\{2,2\}}^{(3)}$.\par

\paragraph{Lagrangian insertion} We will compute the correlation function using the Lagrangian insertion method. For a detailed introduction, we refer to \cite{Alday:2010zy}\cite{Eden:2011yp} and references therein. This method is based on the observation that taking the derivative of the correlation function with respect to the Yang-Mills coupling constant leads to the insertion of one Lagrangian density in the correlation function, \emph{i.e.}
\begin{align}
g^2\frac{\partial}{\partial g^2}G_{\{2,2\}}=\int\rd^4 x_5\langle\mathcal{D}(x_1)\mathcal{D}(x_2)\mathcal{O}_2(x_3)\mathcal{O}_2(x_4)\mathcal{L}(x_5)\rangle
\end{align}
where $\mathcal{L}(x_5)$ is the Lagrangian density and we need to integrate over the position of the inserted operator. Plugging in the perturbative expansion (\ref{eq:pertExpand22}), we find that the $\ell$-loop result can be computed in terms of the Born-level correlation function with $\ell$ Lagrangian insertions. More precisely, we have
\begin{align}
\label{eq:G22l}
G_{\{2,2\}}^{(\ell)}=\int\rd^4 x_5\ldots\rd^4 x_{4+\ell}\,\langle\mathcal{D}(x_1)\ldots\mathcal{O}_2(x_4)
\mathcal{L}(x_5)\ldots\mathcal{L}(x_{4+\ell})\rangle_0
\end{align}
where $\langle\ldots\rangle_0$ denotes the Born level correlation function. To proceed, we need to take two steps. First, we need to compute the Born level $4+\ell$-point correlation function. Second, we need to compute the resulting integrals over $x_5,\ldots,x_{4+\ell}$. It turns out that the first step can be achieved to a large extent by symmetry, which allows us to fix the form of the Born-level correlation function up to a few unknown coefficients. These coefficients can then be fixed by consistency relations and extra input. As for the second step, it turns out the three-loop integrals we encounter are the same as the ones that appeared in the four-point functions of single trace operators. These three-loop integrals have been studied in detail in the literature, see for example \cite{Eden:2011we,Chicherin:2015edu}.

\section{The three-loop integrand}
\label{sec:TQPoly}
In this section, we construct the integrand for $G_{\{2,2\}}^{(\ell)}$ defined in (\ref{eq:G22l}). Let us write
\begin{align}
G_{\{2,2\}}^{(\ell)}(x_1,x_2,x_3,x_4)=\int\rd^4x_5\ldots\rd^4x_{4+\ell}\,\mathcal{G}_{\{2,2\}}^{(\ell)}(x_1,\ldots,x_{4+\ell})\,.
\end{align}
where $\mathcal{G}_{\{2,2\}}^{(\ell)}(x_1,\ldots,x_{4+\ell})$ is the $\ell$-loop \emph{integrand} that we want to compute.

\subsection{Symmetry constraints}
The integrand, or the Born level correlation function $\mathcal{G}_{\{2,2\}}^{(\ell)}(x_1,\ldots,x_{4+\ell})$ is constrained by a number of symmetries.
\paragraph{Supersymmetry} The first constraint comes from supersymmetry. By superconformal Wald identity \cite{Eden:2000bk,Nirschl:2004pa}, the loop correction to the four-point function must be proportional to a \emph{universal factor} defined by
\begin{align}
R_{1234}=&\,d_{12}^2d_{34}^2x_{12}^2x_{34}^2+d_{13}^2d_{24}^2x_{13}^2x_{24}^2+d_{14}^2d_{23}^2x_{14}^2x_{23}^2\\\nonumber
&+d_{12}d_{23}d_{34}d_{14}(x_{13}^2x_{24}^2-x_{12}^2x_{34}^2-x_{14}^2x_{23}^2)\\\nonumber
&+d_{12}d_{13}d_{24}d_{34}(x_{14}^2x_{23}^2-x_{12}^2x_{34}^2-x_{13}^2x_{24}^2)\\\nonumber
&+d_{13}d_{14}d_{23}d_{24}(x_{12}^2x_{34}^2-x_{14}^2x_{23}^2-x_{13}^2x_{24}^2)\,.
\end{align}
where
\begin{align}
d_{ij}\equiv\frac{Y_i\cdot Y_j}{(x_i-x_j)^2},\qquad x_{ij}^2\equiv (x_i-x_j)^2\,.
\end{align}
In what follows, we will also denote $Y_i\cdot Y_j\equiv(y_i-y_j)^2\equiv y_{ij}^2$. We define the conformal cross ratios for $x_j$'s
\begin{align}
u=z\bar{z}=\frac{x_{12}^2x_{34}^2}{x_{13}^2x_{24}^2},\qquad v=(1-z)(1-\bar{z})=\frac{x_{14}^2x_{23}^2}{x_{13}^2x_{24}^2}
\end{align}
and also for the $y_j$'s
\begin{align}
\alpha\bar{\alpha}=\frac{y_{12}^2y_{34}^2}{y_{13}^2y_{24}^2},\qquad (1-\alpha)(1-\bar{\alpha})=\frac{y_{14}^2y_{23}^2}{y_{13}^2y_{24}^2}.
\end{align}
Written in terms of the cross ratios, we have
\begin{align}
R_{1234}=\frac{(z-\alpha)(z-\bar{\alpha})(\bar{z}-\alpha)(\bar{z}-\bar{\alpha})}{z\bar{z}(1-z)(1-\bar{z})}
d_{13}^2d_{24}^2x_{13}^2x_{24}^2\,.
\end{align}

\paragraph{Harmonic and conformal weights} In principle, we can compute the Born-level correlator by Wick contraction. From this perspective, it is clear that in the final result, the $Y_k$'s are contained in the propagators $d_{ij}$ and must take the form of linear combination of polynomials in $y_{ij}^2$. In the computation we must preserve the number of $Y_k$'s. Or, saying in a more fancy way, the harmonic weights. The numbers of $Y_1,Y_2,Y_3,Y_4$ we start with are $N,N,2,2$. The universal quantity $R_{1234}$ contain 2 copies of $Y_k$'s of each type. So we are left with $N-2$ copies of $Y_1$'s and $Y_2$'s and no $Y_3$'s or $Y_4$'s after factorizing out $R_{1234}$. It is easy to see that the only way to encode these $Y_1$'s and $Y_2$'s, under the condition that the result should be a \emph{polynomial} in $Y_k$'s is the factor $(d_{12})^{N-2}$. Therefore we conclude that $\mathcal{G}_{\{2,2\}}^{(\ell)}$ must contain the factor $R_{1234}(d_{12})^{N-2}$ and the rest part is independent of $y_{ij}^2$. Namely, we have
\begin{align}
\mathcal{G}_{\{2,2\}}^{(\ell)}=R_{1234}(d_{12})^{N-2}\,f(x_1,\ldots,x_7)
\end{align}
where $f(x_1,\ldots,x_7)$ only depends on $x_i$.\par 

To constrain the rest part of the integrand, we use conformal symmetry. The conformal weights for the operators $\mathcal{D}(x_i)$ and $\mathcal{O}(x_j)$ are $N/2$ and $1$ respectively. Similar to the four-point function of chiral primary operators, we can further restrict the ansatz by analyzing the OPE structure. For example, in the limit $x_1\to x_2$, the correlator should diverge as $1/(x_{12}^2)^{N}$, which is already taken into account in $R_{1234}(d_{12})^{N-2}$. Similarly, we can consider other the limits $x_i\to x_j$ for $i,j=1,2,3,4$. The divergences from the all such limits should be accounted for in the final result, which poses further constraints for the ansatz.

\paragraph{General ansatz} The above considerations allow us to write down the following ansatz for $\mathcal{G}_{\{2,2\}}^{(\ell)}$
\begin{align}
\label{eq:constructGP}
\mathcal{G}_{\{2,2\}}^{(\ell)}=R_{1234}(d_{12})^{N-2}\frac{P^{(\ell)}}{\prod_{1\le p\le 4\atop 5\le q\le 4+\ell}x_{pq}^2\prod_{5\le p<q\le 4+\ell}x_{pq}^2}
\end{align}
where $P^{(\ell)}$ is a polynomial of $x_{ij}^2$ which carries harmonic weight $0$ and conformal weight $(1-\ell)$ at every point $x_1,\ldots,x_{4+\ell}$. 
As we mentioned in previous section, Lagrangian insertion method enhances extra symmetry of integrands. This new symmetry, called the hidden symmetry in \cite{Eden:2011we}, originates from the crossing symmetry of $n$-point correlation function of operator $\mathcal{T}$, the stress-tensor superfield, and the full permutation symmetry $S_n$ of nilpotent polynomial \cite{Eden:2011we}. 

In our case, these symmetries lead to $S_2 \times S_5$ symmetry
of $P^{(\ell)}$. More explicitly, $P^{(\ell)}$ is symmetric under the permutation of position of \textbf{20}' operators and Lagrangian insertions,  because they belong to the same supermultiplet. These constraints are rather restrictive, especially at low loop orders.

\subsection{Three-loop ansatz}
\begin{figure}[h!]
\centering
\includegraphics[scale=0.4]{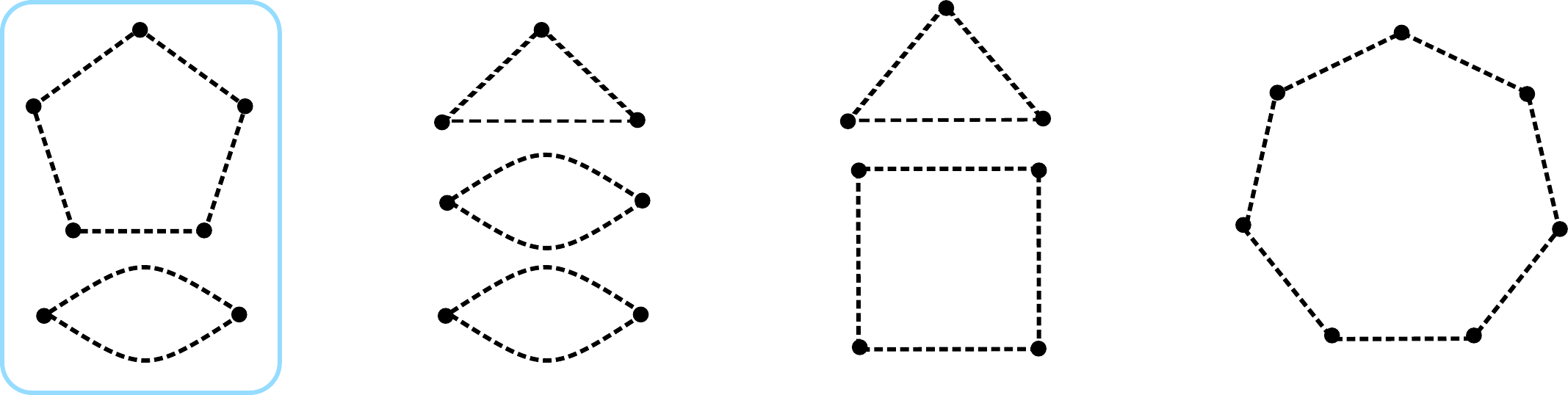}
\caption{Diagrams corresponding to polynomials of degree 7 and weight -2 at each point. Each dot represent one of the coordinates $x_1,\ldots,x_7$. A dashed line between two points, say $x_i$ and $x_j$ represent a factor $x_{ij}^2$. Only the first type of diagram (framed in the blue box) is planar.}
\label{fig:topo}
\end{figure}

Now we focus on the three-loop integrand and construct $P^{(3)}$. From our constraints, $P^{(3)}$ is a homogeneous polynomial of $x_{ij}^2$ of degree 7, with conformal weight $-2$ at each point $x_1,\ldots,x_7$.
It is instructive to represent the polynomial by diagrams. We can represent each spacetime point by a black dot and each $x_{ij}^2$ by a dashed line which connects two dots labeled by $i$ and $j$. In terms of diagrams, our constraints can be rephrased as: given 7 dots, find all the diagrams such that each dot is connected to 2 other dots (required by the conformal weight). It is clear that there are only 4 inequivalent classes of diagrams  as given in figure~\ref{fig:topo}.

For each class in figure~\ref{fig:topo}, we still have different choices of assigning spacetime points to each dots, which corresponds to different monomials. We need to keep in mind that the polynomial $P^{(3)}$ has a $S_2\times S_5$ symmetry, where the $S_2$ is the permutation symmetry of $\mathcal{D}(x_1)$ and $\mathcal{D}(x_2)$ and $S_5$ is the permutation symmetry for the single-trace operators and the Lagrangian densities.\par

\paragraph{Planarity} Planarity imposes further constraints for the type of diagrams that we need to consider. The role of planarity has been discussed in detail in \cite{Chicherin:2015edu}. Here we impose the same planarity criteria as in \cite{Chicherin:2015edu}. Namely, we consider planarity of the quantity $(d_{12})^{N-2}f^{(\ell)}(x_1,\ldots,x_7)$ where
\begin{align}
f^{(\ell)}(x_1,\ldots,x_7)=\frac{P^{(\ell)}(x_1,\ldots,x_7)}{\prod_{1\le p\le 4\atop 5\le q\le 4+\ell}x_{pq}^2\prod_{5\le p<q\le 4+\ell}x_{pq}^2}
\end{align}
in the following way. We draw a solid line between two points $x_i$ and $x_j$ if there is a factor $1/(x_{ij}^2)$ and a dashed line if there is factor $x_{ij}^2$ in $f^{(\ell)}(x_1,\ldots,x_7)$. A solid line and a dashed line connecting two points cancel each other. If all solid lines can be drawn on a sphere, we call the corresponding quantity `planar', otherwise it is non-planar. The main assumption we made here is that, in computing the four-point function in the planar limit, we only need to take into account the polynomials $P^{(\ell)}$ such that $f^{(\ell)}$ is planar. Under this assumption, one finds that only the first class in figure~\ref{fig:topo} contributes.

\paragraph{Final ansatz}
\begin{figure}[h!]
\centering
\includegraphics[scale=0.4]{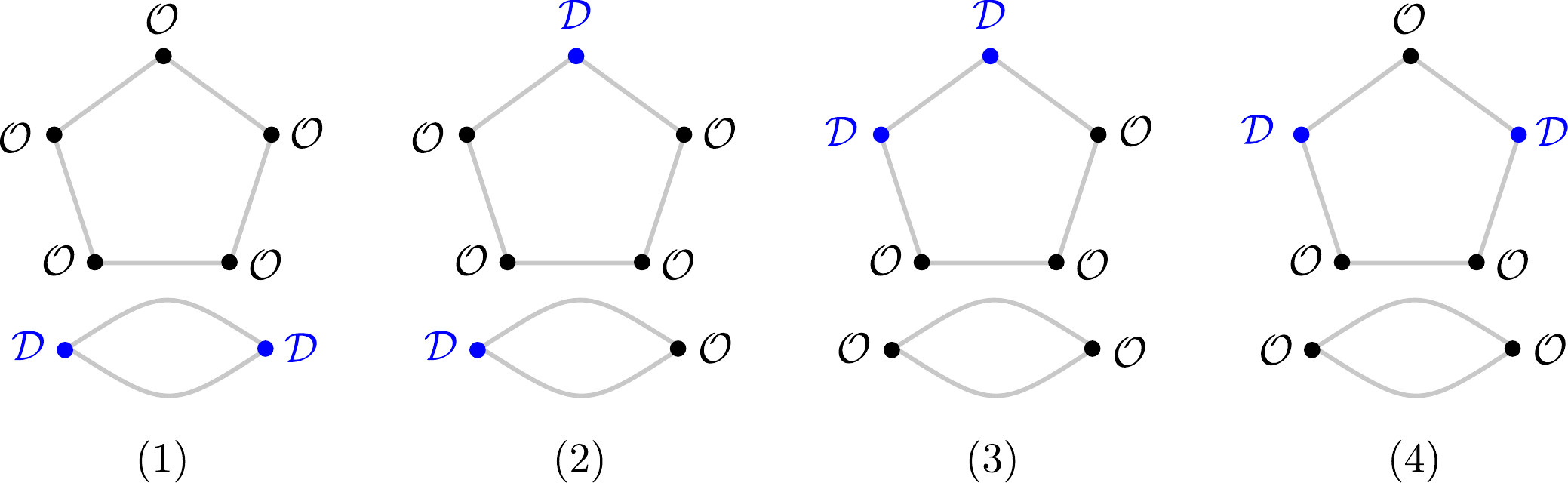}
\caption{Diagrams for the integrand in the planar limit. The blue and black dots denote giant gravitons and single-trace operators, respectively. For each diagram, we need to take into account the permutations of $x_1,x_2$ for the giant graviton and $x_3,\ldots,x_7$ of the single-trace operators.}
\label{fig:diagrams}
\end{figure}
After imposing planarity, we still need to distinguish between giant gravitons and the single-trace operators. Taking into account the $S_2\times S_5$ symmetry, we find four in-equivalent diagrams as is shown in figure~\ref{fig:diagrams}. The four diagrams correspond to the following polynomials
\begin{align}
\label{eq:P3j}
&P_1^{(3)}(x_i)=(x_{12}^4)(x_{34}^2x_{45}^2x_{56}^2x_{67}^2x_{73}^2)+S_2\times S_5\text{ permutations}\,,\\\nonumber
&P_2^{(3)}(x_i)=(x_{13}^4)(x_{24}^2x_{45}^2x_{56}^2x_{67}^2x_{72}^2)+S_2\times S_5\text{ permutations}\,\\\nonumber
&P_3^{(3)}(x_i)=(x_{67}^4)(x_{12}^2x_{23}^2x_{34}^2x_{45}^2x_{51}^2)+S_2\times S_5\text{ permutations}\,,\\\nonumber
&P_4^{(3)}(x_i)=(x_{67}^4)(x_{13}^2x_{32}^2x_{24}^2x_{45}^2x_{51}^2)+S_2\times S_5\text{ permutations}\,.
\end{align}
The most general ansatz for the integrand is the linear combination of the above four polynomials
\begin{align}
P^{(3)}(x_i)=c_1\,P_1^{(3)}(x_i)+c_2\,P_2^{(3)}(x_i)+c_3\,P_3^{(3)}(x_i)+c_4\,P_4^{(3)}(x_i)
\end{align}
where $c_1,\ldots,c_4$ are coefficients to be fixed.

\subsection{The three-loop integral} 
To compute the correlation function we plug $P_j^{(3)}$ \eqref{eq:P3j} into \eqref{eq:constructGP} and perform the integrals over $x_5,x_6,x_7$. It turns out that the resulting integrals can all be written in terms of the three-loop integrals that have appeared in the computation of the four-point function of \textbf{20}' operators. The definition of these integrals are summarized in appendix~\ref{app:integrals}. Let us denote the integrals
\begin{align}
 I_j^{(3)}\equiv x_{13}^2x_{24}^2\int \frac{P_j^{(3)}(x_1,\ldots,x_7)}{\prod_{1\le p\le 4\atop 5\le q\le 4+\ell}x_{pq}^2\prod_{5\le p<q\le 4+\ell}x_{pq}^2}\,\mathrm{d}^4 x_5\mathrm{d}^4x_6\mathrm{d}^4x_7\,.
\end{align}
The four integrals can be written in terms of the known three-loop integrals as follows
\begin{align}
I_1^{(3)}=&\,6\left[gh(2,4;1,3)+gh(2,3;1,4)+gh(1,4;2,3)+gh(1,3;2,4)\right]\\\nonumber
&\,+6\left[L(2,4;1,3)+L(1,4;2,3)+L(2,3;1,4)+L(1,3;2,4)\right]\\\nonumber
&\,+6\left[E(1,2;3,4)+E(2,1;3,4)+E(1,3;2,4)+E(2,3;1,4)+E(1,4;2,3)+E(2,4;1,3)\right]\\\nonumber
&\,+6\left[T(1,4;2,3)+T(2,4;1,3)+T(1,3;2,4)+T(2,3;1,4)+T(1,2;3,4)+T(2,1;3,4)\right]\,,\\\nonumber
I_2^{(3)}=&\,6\left[gh(3,4;1,2)+L(3,4;1,2) \right]\,,\\\nonumber
I_3^{(3)}=&\,6\left[gh(1,2;3,4)+T(4,3;1,2)+T(3,4;1,2)\right]\\\nonumber
&\,+6\left[E(4,1;2,3)+E(3,1;2,4)+E(4,2;1,3)+E(3,2;1,4)\right]\\\nonumber
&\,+3\left[H(1,4;3,2)+H(2,4;3,1)+H(1,3;4,2)+H(2,3;4,1) \right]\\\nonumber
&\,+3\left[H(3,4;1,2)+H(3,4;2,1)\right]\,,\\\nonumber
I_4^{(3)}=&\,6\left[L(1,2;3,4)+T(4,1;2,3)+T(4,2;1,3)\right]\\\nonumber
&\,+6\left[T(3,1;2,4)+T(3,2;1,4)+E(4,3;1,2)+E(3,4;1,2)\right]\\\nonumber
&\,+3\left[H(1,2;3,4)+H(1,2;4,3)+H(1,4;2,3)\right]\\\nonumber
&\,+3\left[H(2,4;1,3)+H(1,3;2,4)+H(2,3;1,4)\right]\,.
\end{align}
Our final result is given by the linear combination of these integrals
\begin{align}
\label{eq:bigAnsatz}
\frac{G_{\{2,2\}}^{(3)}}{\tilde{R}_{1234}(d_{12})^{N-2}}=c_1\,I_1^{(3)}+c_2\,I_2^{(3)}+c_3\,I_3^{(3)}+c_4\,I_4^{(3)}\,
\end{align}
where $\tilde{R}_{1234}=R_{1234}/(x_{13}^2x_{24}^2)$. To fix the four unknown coefficients, we need extra input. In the case of single-trace operators, one powerful constraint comes from the light-like limit where one can apply the duality between correlation functions and amplitudes \cite{Alday:2010zy,Eden:2010zz}. In the presence of the giant gravitons, we are not aware of such dualities. Therefore we need other means to fix these coefficients. In the next section we shall consider the OPE limit of the four-point functions in two different channels, which allows us to fix these coefficients. After obtaining the explicit result, we can then take the light-like limit of the four-point function. We find that, interestingly, the result (normalized by the Born level four-point function) is again the 3-loop four-point gluon MHV amplitude. This hints that the correlator/amplitude duality still holds in the presence of giant gravitons. For more details and discussions, we refer to section~\ref{eq:subsecLL}.

\section{OPE limit}
\label{sec:OPElimit}
\begin{figure}[h!]
\centering
\includegraphics[scale=0.5]{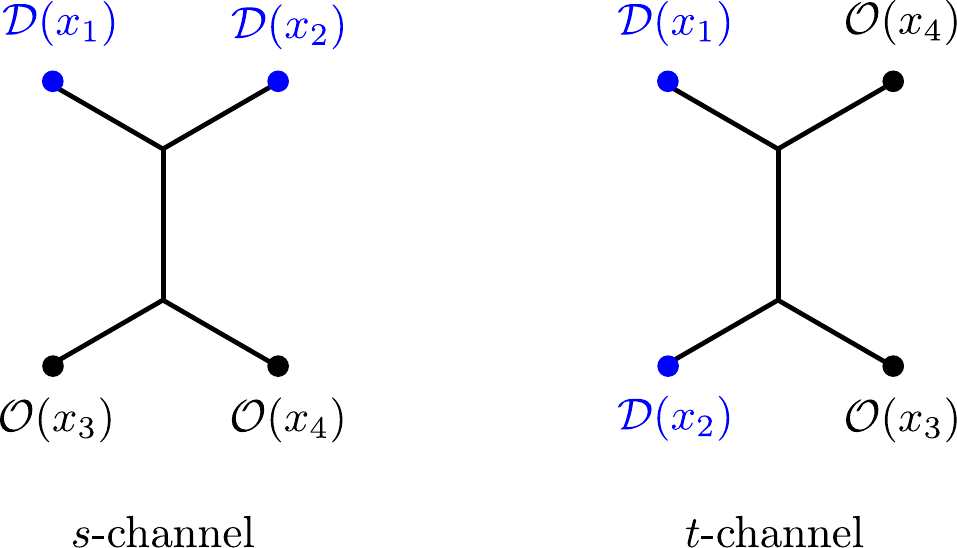}
\caption{OPE in two different channels.}
\label{fig:OPE}
\end{figure}

In the OPE limit, the leading contributions are controlled by a few conformal data such as the anomalous dimensions and OPE coefficients of low lying operators. It turns out to fix the four coefficients of the three-loop ansatz, it is sufficient to plug in two-loop conformal data for the low lying operators, which are already known in the literature. Because we have two kinds of operators $\mathcal{D}(x_i)$ and $\mathcal{O}(x_j)$, we also have two different OPE channels, which we shall call the $s$- and $t$-channel OPEs, as is shown in figure~\ref{fig:OPE}.

\subsection{OPE in the $s$-channel}
The $s$-channel OPE limit is defined by $x_1\to x_2$ and $x_3 \to x_4$. In terms of conformal cross ratios, this corresponds to $u\rightarrow 0$ and $v \rightarrow 1$. In this limit, we replace the product $\mathcal{O}(x_3)\mathcal{O}(x_4)$ by the following series of operators,
\begin{align}
\label{eq:OPEoo}
\mathcal{O}(x_3)\mathcal{O}(x_4)=&\,\mathsf{c}_{\mathcal{I}}\frac{(Y_3 \cdot Y_4)^2}{(x_{34}^2)^2}\mathcal{I}+\mathsf{c}_{\mathcal{O}}\frac{(Y_3 \cdot Y_4)}{x_{34}^2}Y_{3I}Y_{4J}\mathcal{O}_{\mathbf{20'}}^{IJ}(x_4)\\\nonumber
&\,+\mathsf{c}_{\mathcal{K}}(g)\frac{(Y_3 \cdot Y_4)^2}{(x_{34}^2)^{1-\delta \Delta_{\mathcal{K}}/2}}\mathcal{K}(x_4)+\ldots,
\end{align}
where $\mathcal{I}$ is the identity operator,  $\mathcal{K}$ is the Konishi operator, and $\mathcal{O}^{IJ}_{\mathbf{20'}}$ belongs to the $\mathbf{20'}$ representation of $SU(4)$. The conformal data of these operators control the leading and subleading contributions in the OPE limit $x_3\to x_4$. Since $\mathcal{O}_{\mathbf{20'}}^{IJ}$ is a half-BPS operator, its anomalous dimension vanishes and $c_{\mathcal{O}}$ doesn't depend on the coupling. Therefore the only coupling dependence comes from anomalous dimension and OPE coefficient of Konishi operator.\par 

Similarly, we can perform the OPE of $\mathcal{D}(x_1) \mathcal{D}(x_2)$ 
\begin{align}
\label{eq:OPEdd}
\mathcal{D}(x_1)\mathcal{D}(x_2)=&\,\mathsf{d}_{\mathcal{I}}\frac{(Y_1\cdot Y_2)^{N}}{(x_{12}^2)^{N}}
+\mathsf{d}_{\mathcal{O}}\frac{(Y_1\cdot Y_2)^{N-1}}{(x_{12}^2)^{N-1}}Y_{3I}Y_{4J}\mathcal{O}_{\mathbf{20'}}^{IJ}(x_4)\\\nonumber
&\,+\mathsf{d}_{\mathcal{K}}(g)\frac{(Y_1\cdot Y_2)^{N}}{(x_{12}^2)^{N-1-\delta\Delta_{\mathcal{K}}/2}}\mathcal{K}(x_2)+\ldots
\end{align}
Again the coupling dependence comes from the OPE data of Konish operator. We insert the OPEs \eqref{eq:OPEoo} and \eqref{eq:OPEdd} into the four-point function and expand up to three-loop order, leading to
\begin{align}
\label{eq:3loopG}
\lim_{x_1\to x_2\atop x_3\rightarrow x_4}G_{\{2,2\}}^{(3)}=\left.\mathsf{d}_{\mathcal{K}}(g)\mathsf{c}_{\mathcal{K}}(g)\frac{(Y_1\cdot Y_2)^{N}(Y_3 \cdot Y_4)^2}{(x_{12}^2)^{N-1-\delta\Delta_{\mathcal{K}}/2}(x_{34}^2)^{1-\delta \Delta_{\mathcal{K}}/2}}\left\langle \mathcal{K}(x_2)\mathcal{K}(x_4)\right\rangle \right|_{O(g^6)}+\cdots\,,
\end{align}
where we normalize the two-point function of Konishi operator to be 
\begin{align}
\label{eq:KK2pt}
\left\langle \mathcal{K}(x_2)\mathcal{K}(x_4)\right\rangle=\frac{1}{(x_{24}^2)^{2+\delta \Delta_{\mathcal{K}}}}\,.
\end{align}
Plugging \eqref{eq:KK2pt} into \eqref{eq:3loopG}, we find that
\begin{align}
\label{eq:G322O3}
\lim_{x_1\to x_2\atop x_3\rightarrow x_4}G_{\{2,2\}}^{(3)}=\left.\frac{y_{12}^4 y_{34}^4}{x_{12}^2 x_{34}^2 x_{24}^4}(d_{12})^{N-2}\mathsf{d}_{\mathcal{K}}(g)\mathsf{c}_{\mathcal{K}}(g){u^{\delta \Delta_{\mathcal{K}}/2}}\right|_{O(g^6)}\,.
\end{align}
The anomalous dimension of the Konishi operator and OPE coefficients can be expanded perturbatively
\begin{align}
\label{eq:dataSchannel}
\delta\Delta_{\mathcal{K}}(g)=&\,g^2\,\gamma_1+g^4\,\gamma_2+g^6\,\gamma_3+\cdots, \\\nonumber
\mathsf{d}_{\mathcal{K}}(g)\mathsf{c}_{\mathcal{K}}(g)=&\,C_0+g^2\,C_1+g^4\,C_2+g^6\,C_3+\cdots\,.
\end{align}
The coupling dependent part \eqref{eq:G322O3} thus can be expanded as
\begin{align}
\label{eq:expanddcu}
\left.\mathsf{d}_{\mathcal{K}}(g)\mathsf{c}_{\mathcal{K}}(g){u^{\delta \Delta_{\mathcal{K}}/2}}\right|_{O(g^6)}=&\,\frac{1}{48}C_0\,\gamma_1^3\,(\ln u)^3
+\frac{1}{8}\left(C_1\,\gamma_1^2+2C_0\,\gamma_1\gamma_2 \right)(\ln u)^2\\\nonumber
&\,+\frac{1}{2}\left(C_0\,\gamma_3+C_1\,\gamma_2+C_2\,\gamma_1\right)(\ln u)+C_3\,.
\end{align}
We notice that the result can be organized as a polynomial in $\ln u$. This structure plays an important role in our analysis below. In the expansion \eqref{eq:expanddcu}, the anomalous dimension of the Konishi operator and other twist-2 operators can be computed effectively by quantum spectral curve \cite{Gromov:2013pga} up to very high orders (see for example \cite{Marboe:2014sya} and references therein), so $\gamma_n$ is known. In addition, the coefficients $C_k$ contain the perturbative expansions of both $\mathsf{d}_{\mathcal{K}}(g)$ and $\mathsf{c}_{\mathcal{K}}(g)$. The result for $\mathsf{c}_{\mathcal{K}}(g)$ can be computed by the hexagon approach \cite{Basso:2015zoa} and is known and checked up to 5-loop order \cite{Basso:2022nny,Georgoudis:2017meq}. The perturbative expansion of $\mathsf{d}_{\mathcal{K}}(g)$ can be computed by the $g$-function approach and has been tested up to two-loop order. To sum up, apart from $C_3$, we know all the other coefficients in the expansion \eqref{eq:expanddcu}.

\paragraph{$s$-channel OPE from the 3-loop ansatz} Now let us take the $s$-channel OPE limit of the 3-loop ansatz \eqref{eq:bigAnsatz}. We can use the symmetry of the integrals listed in appendix~\ref{app:integrals} to simplify the ansatz \eqref{eq:bigAnsatz} and write the results in terms of the three-loop conformal integrals
\begin{align}
\label{eq:simplifyG}
\frac{G_{\{2,2\}}^{(3)}}{\tilde{R}_{1234}(d_{12})^{N-2}}=&\,12(c_1+c_4){E(1,2;3,4)}+12(c_1+c_3)[{E(1,3;2,4)}+{E(1,4;2,3)}]\\\nonumber
&\,+12(2c_1+c_2+2c_3+c_4){L(1,2;3,4)}\\\nonumber
&\,+12(2c_1+c_4)[{L(1,3;2,4)}+{L(1,4;2,3)}]\\\nonumber
&\,+6(c_2+c_3){gh(1,2;3,4)}+12c_1[{gh(1,3;2,4)}+{gh(1,4;2,3)}]\\\nonumber
&\,+3(c_3+c_4)(1+1/v){H(1,2;3,4)}+6(c_4+c_3\,u/v){H(1,3;2,4)}\\\nonumber
&\,+6(c_4+c_3\,u){H(1,4;2,3)}
\end{align}
The result for the integrals \eqref{eq:simplifyG} in the $s$-channel OPE limit can be computed and are given in appendix~\ref{app:integrals}. Substitute the $s$-channel asymptotic results in \eqref{eq:simplifyG}, we obtain
\begin{align}
\label{eq:G22O3OPEs}
\frac{G_{\{2,2\}}^{(3)}}{\tilde{R}_{1234}(d_{12})^{N-2}}=&\,\frac{3}{16}(2c_1+c_4)(\ln u)^3-\frac{3}{16}(19c_1+11c_4)(\ln u)^2\\\nonumber
&\,+\frac{3}{8}\left[(37c_1+24c_4)+3(2c_1+3c_3+c_4)\zeta_3\right]\ln u\\\nonumber
&\,-\frac{9}{8}(19c_1+13c_4)-\frac{3}{8}(9c_1+15c_3-8c_4)\zeta_3\\\nonumber
&\,-\frac{15}{8}(4c_1+c_2+2c_3+3c_4)\zeta_5
\end{align}
We can now compare \eqref{eq:G322O3} and \eqref{eq:G22O3OPEs}. As we already discussed before, the coefficients of $(\ln u)^n$ with $n\ge 1$ contain only lower loop data. The constant term is our 3-loop prediction. In this way, we obtain 3 equations, corresponding to the 3 coefficients in front of $(\ln u)^n$ with $n=1,2,3$:
\begin{align}
24c_1+12c_4=&\,\frac{1}{48}C_0\gamma_1^3\,,\\\nonumber
-228c_1-132c_4=&\,\frac{1}{8}(C_1\gamma_1^2+2C_0\gamma_1\gamma_2)\,,\\\nonumber
888c_1+576c_4+72(2c_1+3c_2+c_4)\zeta(3)=&\,\frac{1}{2}\left(C_0\gamma_3+C_1\gamma_2+C_2\gamma_1\right)\,.
\end{align}
Plugging in the following perturbative data up to two-loops
\begin{align}
\delta \Delta_{\mathcal{K}}(g)=&\,\gamma_1\,g^2+\gamma_2\,g^4+\gamma_3\,g^6+\cdots=12g^2-48g^4+336g^6+\cdots\,,\\\nonumber
\mathsf{d}_{\mathcal{K}}(g)\mathsf{c}_{\mathcal{K}}(g)=&\,C_0+C_1\,g^2+C_2\,g^4+\cdots=\frac{1}{3}-4g^2+56g^4+\cdots\,,
\end{align}
we find the following simple solution
\begin{align}
\label{eq:schannelsolution}
c_1=\frac{1}{3},\qquad c_2=-\frac{1}{3},\qquad c_4=\frac{1}{3}\,.
\end{align}
Therefore the $s$-channel OPE alone fixes 3 out of 4 unknown coefficients. To obtain $c_3$, we need to consider the $t$-channel OPE.

\subsection{OPE in the $t$-channel}
Now we consider the $t$-channel OPE limit $x_1\rightarrow x_4$ and $x_2 \rightarrow x_3$, which is equivalently to $u\rightarrow 1$ and $v \rightarrow 0$. In this channel, the dominant contribution comes from an operator with dimension $N+\delta\Delta_{\text{open}}(g)$ where $N$ is the bare dimension. We have the following OPE
\begin{align}
\label{eq:OPEtchannel}
\mathcal{D}(x_1)\mathcal{O}(x_4)=\mathsf{c}_{\mathcal{O}\text{open}}(g)\frac{(Y_1\cdot Y_4)^2}{(x_{14}^2)^{1-\delta\Delta_{\text{open}}/2}}\mathcal{O}_{\text{open}}+\cdots
\end{align}
To have the correct harmonic charge and bare dimension, the candidate operator should take the form
\begin{equation}
	\mathcal{O}_{\text{open}}\sim\epsilon^{j_1 ...j_{N-2}a_1a_2}_{i_1...i_{N-2}b_1b_2}(\mathcal{Z}_1)^{i_1}_{j_1}\cdots(\mathcal{Z}_1)^{i_{N-2}}_{j_{N-2}}(\Phi_I)_{b_1}^{a_1}(\Phi_I)_{b_2}^{a_2}.
\end{equation}
where $\mathcal{Z}_1=Y_1^I\Phi_I$ and we sum over $I=1,\ldots,6$ for the last two scalar fields. This is not a familiar operator of the $Z=0$ or $Y=0$ brane excitations. In \cite{Jiang:2019xdz} it was conjectured that $\mathcal{O}_{\text{open}}$ corresponds to the $Z=0$ brane with length-1 excitation. This does not seem to be the case. However, in fact we do not have to make further assumptions about this operator, the CFT data $\mathsf{c}_{\mathcal{O}\text{open}}(g)$ and $\delta\Delta_{\text{open}}(g)$ can be fixed up to two-loop order solely from lower loop data. We give the main idea here and the details can be found in appendix~\ref{app:pert2loop}. Let us denote the perturbative expansion of the OPE data as
\begin{align}
(\mathsf{c}_{\mathcal{O}\text{open}}(g))^2=&\,\widetilde{C}_0+\widetilde{C}_1\,g^2+\widetilde{C}_2\,g^4+\widetilde{C}_3\,g^6+\cdots\\\nonumber
\delta\Delta_{\text{open}}(g)=&\,\tilde{\gamma}_1\,g^2+\tilde{\gamma}_2\,g^4+\tilde{\gamma}_3\,g^6+\cdots
\end{align}
where $\widetilde{C}_0$ can be fixed easily from a Born level computation. At one-loop, there is only one unknown coefficient, which can be fixed by the $s$-channel OPE. This gives us the full one-loop answer. By performing a $t$-channel OPE, we can extract $\mathsf{c}_{\mathcal{O}\text{open}}(g)$ and $\delta\Delta_{\text{open}}(g)$ up to one-loop. At two-loop, there are two unknown coefficients, one of which can be fixed by the $s$-channel OPE. The other one can be fixed by $t$-channel OPE. It turns out that there are several equations in the $t$-channel OPE, one of them involve only lower order OPE data $\widetilde{C}_0$, $\widetilde{C}_1$ and $\tilde{\gamma}_1$, which have been determined in the previous step. We can use this equation to fix the second unknown coefficient, which then give us the full two-loop result. Performing the $t$-channel OPE, we can extract $\widetilde{C}_2$ and $\tilde{\gamma}_2$. This procedure leads to the following result
\begin{align}
\label{eq:tchanneldata}
(\mathsf{c}_{\mathcal{O}\text{open}}(g))^2=&\,1-4g^2+24(1+\zeta_3)g^4+\cdots,\\\nonumber
\delta\Delta_{\text{open}}(g)=&\,4g^2-8g^4+\cdots\,.
\end{align}
Using the $t$-channel OPE \eqref{eq:OPEtchannel} and a similar one for $\mathcal{D}(x_2)$ and $\mathcal{O}(x_3)$, we obtain the correlator in the OPE limit
\begin{align}
\lim_{x_4\to x_1\atop x_3\to x_2} G_{\{2,2\}}^{(3)}=\left.{(\mathsf{c}_{\mathcal{O}\text{open}})^2}\frac{(Y_1\cdot Y_4)^2(Y_2\cdot Y_3)^2}{(x_{14}^2)^{1-\delta\Delta_{\text{open}}/2}(x_{23}^2)^{1-\delta\Delta_{\text{open}}/2}}
\langle\mathcal{O}_{\text{open}}(x_1)\mathcal{O}_{\text{open}}(x_2)\rangle\right|_{{O}(g^6)}+\cdots
\end{align}
Using 
\begin{align}
\langle\mathcal{O}_{\text{open}}(x_1)\mathcal{O}_{\text{open}}(x_2)\rangle\propto\frac{d_{12}^{N-2}}{(x_{12}^2)^{2+\delta\Delta_{\text{open}}}}\,,
\end{align}
we obtain
\begin{align}
\lim_{x_4\to x_1\atop x_3\to x_2} G_{\{2,2\}}^{(3)}=\left.{(\mathsf{c}_{\mathcal{O}\text{open}})^2}\frac{y_{14}^4 y_{23}^4}{x_{14}^2x_{23}^2x_{12}^4}v^{\delta\Delta_{\text{open}}/2}\right|_{{O}(g^6)}
\end{align}
We have the following perturbative expansion
\begin{align}
\label{eq:expandcopen}
\left.\left(\mathsf{c}_{\mathcal{O}\text{open}}(g)\right)^2{v^{\delta \Delta_{\text{open}}/2}}\right|_{O(g^6)}=&\,\frac{1}{48}\widetilde{C}_0\,\tilde{\gamma}_1^3\,(\ln v)^3
+\frac{1}{8}\left(\widetilde{C}_1\,\tilde{\gamma}_1^2+2\widetilde{C}_0\,\tilde{\gamma}_1\tilde{\gamma}_2 \right)(\ln v)^2\\\nonumber
&\,+\frac{1}{2}\left(\widetilde{C}_0\,\tilde{\gamma}_3+\widetilde{C}_1\,\tilde{\gamma}_2+\widetilde{C}_2\,\tilde{\gamma}_1\right)(\ln v)+\widetilde{C}_3\,.
\end{align}
On the other hand, we can take the $t$-channel OPE limit in our ansatz \eqref{eq:simplifyG} using the results of the three-loop integrals in the $t$-channel OPE limit, which leads to
\begin{align}
\lim_{x_4\to x_1\atop x_3\to x_2}\frac{G_{\{2,2\}}^{(3)}}{\tilde{R}_{1234}(d_{12})^{N-2}}=&\,\frac{1}{16}(4c_1+c_2+3c_3+c_4)(\ln v)^3\\\nonumber
&\,-\frac{3}{16}(13c_1+3c_2+10c_3+4c_4)(\ln v)^2\\\nonumber
&\,+\frac{3}{32}\left[(104c_1+22c_2+79c_3+39c_4)+24(c_1+c_3+c_4)\zeta(3)\right](\ln v)\\\nonumber
&\,-\frac{3}{16}(82c_1+16c_2+57c_3+37c_4)-\frac{3}{8}(9c_1+7c_4)\zeta(3)\\\nonumber
&\,-\frac{15}{4}(3c_1+c_3+c_4)\zeta(5)
\end{align}
Collecting the coefficients of $(\ln v)^n$ with $n\ge 1$ and comparing with \eqref{eq:expandcopen}, we obtain the following equations
\begin{align}
\label{eq:tchannelEq}
16c_1+4c_2+12c_3+4c_4=&\,\frac{1}{48}\widetilde{C}_0\,\tilde{\gamma}_1^3\,,\\\nonumber
-156c_1-36c_2-120c_3-48c_4=&\,\frac{1}{8}\left(\widetilde{C}_1\,\tilde{\gamma}_1^2+2\widetilde{C}_0\,\tilde{\gamma}_1\tilde{\gamma}_2 \right)\,,\\\nonumber
6(104 c_1+22c_2+79c_3+39c_4)+144(c_1+c_3+c_4)\zeta_3=&\,\frac{1}{2}\left(\widetilde{C}_0\,\tilde{\gamma}_3+\widetilde{C}_1\,\tilde{\gamma}_2+\widetilde{C}_2\,\tilde{\gamma}_1\right)\,.
\end{align}
Since we only need to determine $c_3$, it is sufficient to consider the first equation in \eqref{eq:tchannelEq}. Plugging in the perturbative data \eqref{eq:tchanneldata} and the solution from $s$-channel \eqref{eq:schannelsolution}, we find that
\begin{align}
c_3=-\frac{1}{3}\,.
\end{align}
We can then use the second equation of \eqref{eq:tchannelEq} as a consistency check, which is indeed satisfied. Plugging the solutions into \eqref{eq:simplifyG}, we obtain
\begin{align}
\label{eq:finalRes}
\frac{G^{(3)}_{\{2,2\}}}{\tilde{R}_{1234}(d_{12})^{N-2}}=&\,4\left[gh(1,3;2,4)+gh(1,4;2,3)-gh(1,2;3,4)\right]\\\nonumber
&\,+12\left[L(1,3;2,4)+L(1,4;2,3)\right]+8 E(1,2;3,4)\\\nonumber
&\,+2\left(1-\frac{u}{v}\right)H(1,3;2,4)+2(1-u)H(1,4;2,3)\,.
\end{align}
This is the main result of the current work. We note that it is similar to the three-loop result of four single-trace operators, but somewhat simpler. Not all the three-loop integrals appear.

\subsection{Light-like limit}
\label{eq:subsecLL}
In this subsection, we consider the light-like limit $x_{i,i+1}^2\to 0$ of the four-point function $G_{\{2,2\}}/G_{\{2,2\}}^{(0)}$ with $n=1,2,3$. In the case of four BPS single-trace operators of length-2, this quantity corresponds to the square of the four-gluon MHV amplitude. This fact was called the correlator/amplitude duality \cite{Alday:2010zy,Eden:2010zz}. We shall find that this duality still holds in the presence of giant gravitons.
\paragraph{Tree level} The tree-level result reads \cite{Jiang:2019xdz,Jiang:2019zig}
\begin{align}
\frac{G_{\{2,2\}}^{(0)}}{(d_{12})^N(d_{34})^2}=\frac{z\bar{z}}{\alpha\bar{\alpha}}\left[\frac{(1-\alpha)(1-\bar{\alpha})}{(1-z)(1-\bar{z})}+1 \right]-2\left(\frac{z\bar{z}}{\alpha\bar{\alpha}}\right)^2\frac{(1-\alpha)(1-\bar{\alpha})}{(1-z)(1-\bar{z})}\,.
\end{align}
In the light-like limit $x_{i,i+1}^2\to 0$, we have
\begin{align}
z\bar{z}=\frac{x_{12}^2x_{34}^2}{x_{13}^2x_{24}^2}\to 0,\qquad (1-z)(1-\bar{z})=\frac{x_{14}^2x_{23}^2}{x_{13}^2x_{24}^2}\to 0
\end{align}
Keeping the leading term in this limit, we find that
\begin{align}
\label{eq:G0ll}
G_{\{2,2\}}^{(0)}\sim (d_{12})^N(d_{34})^2\frac{z\bar{z}}{\alpha\bar{\alpha}}\frac{(1-\alpha)(1-\bar{\alpha})}{(1-z)(1-\bar{z})}
=\frac{x_{12}^2x_{34}^2}{x_{14}^2x_{23}^2}\frac{y_{14}^2y_{23}^2}{y_{12}^2y_{34}^2}\,(d_{12})^N(d_{34})^2
\end{align}
At loop level, the global factor $\tilde{R}_{1234}(d_{12})^{N-2}$ in the light-like limit becomes
\begin{align}
\frac{R_{1234}}{x_{13}^2x_{24}^2}(d_{12})^{N-2}\sim \frac{y_{12}^2y_{23}^2y_{34}^2y_{14}^2}{x_{12}^2x_{23}^2x_{34}^2x_{14}^2}(d_{12})^{N-2}=\frac{y_{23}^2y_{14}^2}{x_{23}^2x_{14}^2}\frac{x_{12}^2x_{34}^2}{y_{12}^2y_{34}^2}(d_{12})^N(d_{34})^2\,.
\end{align}
which is cancelled precisely by dividing $G_{\{2,2\}}^{(0)}$.
\paragraph{One-loop} At one-loop level, we have
\begin{align}
\lim_{x_{i,i+1}^2\to 0}\frac{G_{\{2,2\}}^{(1)}}{G_{\{2,2\}}^{(0)}}=2 g(1,2,3,4)\,.
\end{align}

\paragraph{Two-loop} At two-loop, we have
\begin{align}
\lim_{x_{i,i+1}^2\to 0}\frac{G_{\{2,2\}}^{(2)}}{G_{\{2,2\}}^{(0)}}=2h(1,3;2,4)+2h(2,4;1,3)+g(1,2,3,4)^2
\end{align}

\paragraph{Three-loop} At three-loop, we have
\begin{align}
\lim_{x_{i,i+1}^2\to 0}\frac{G_{\{2,2\}}^{(3)}}{G_{\{2,2\}}^{(0)}}=&\,2gh(1,3;2,4)+2gh(2,4;1,3)\\\nonumber
&\,+2L(2,4;1,3)+2L(1,3;2,4)+2T(2,4;1,3)\\\nonumber
&\,+2T(1,3;2,4)+2T(4,2;1,3)+2T(3,1;2,4)\,.
\end{align}
Noticing that in taking the light-like limit, some of the integral identities do not hold anymore. For example, the $T$-integrals do not equal to $L$-integrals in this limit. On the other hand, the four-gluon MHV amplitude up to three-loop order is given by\cite{Drummond:2006rz}\cite{Anastasiou:2003kj}
\begin{align}
\frac{\mathcal{A}}{\mathcal{A}^{(0)}}=&\,1+g^2\,[g(1,2,3,4]+g^4[h(1,3;2,4)+h(2,4;1,3)]\\\nonumber
&\,+g^6[T(1,3;2,4)+T(3,1;2,4)+T(2,4;1,3)+T(4,2;1,3)\\\nonumber
&\,\quad+L(1,3;2,4)+L(2,4;1,3)]\,.
\end{align}
From these explicit results, we find that up to three-loop order
\begin{align}
\lim_{x_{i,i+1}^2\to 0}\frac{G_{\{2,2\}}}{G^{(0)}_{\{2,2\}}}=\left(\frac{\mathcal{A}}{\mathcal{A}^{(0)}}\right)^2+\mathcal{O}(g^8)\,.
\end{align}
Therefore in the light-like limit, the four-point function is also the square of the MHV amplitude. It is natural to conjecture that this relation holds at higher-loop orders, or even non-perturbatively.\par

If this were the case, we can use this relation to constrain the four-point functions with giant gravitons. However, we would like to point out that, this is only sufficient to fix part of the unknown coefficients both at two-loop and three-loops. These are the coefficients that can be fixed by the $s$-channel OPE limit. This is different from the case where the four-operators are length-2 BPS operators. The reason is that the latter has a larger permutation symmetry since the four single-trace operators are identical. Therefore we have fewer unknown coefficients, which can be fixed completely by the correlator/amplitude duality.

\section{OPE data at three-loop order}
\label{sec:OPE}
From the explicit form of the four-point function, we can extract the OPE data up to this loop order. For length-2 single-trace BPS operators, at the leading OPE limit in the $s$-channel we can extract OPE data for twist-2 operators. There is no degeneracy for twist-2 operators and we have one operator for each spin. We denote the twist-2 operator with spin-$S$ by $\mathcal{O}_S$. The OPE coefficients of two giant gravitons and two length-$L$ half-BPS operators will be denoted by $\mathsf{d}_{S}$ and $\mathsf{c}_{LLS}$ respectively.

From our result \eqref{eq:finalRes}, we can extract the product of OPE coefficients $\mathsf{d}_{S}\mathsf{c}_{22S}$ at three-loop order using the method described in \cite{Vieira:2013wya}. The result is given in Table~\ref{tab:dataDC}.
\begin{table}[h!]
\begin{center}
\renewcommand\arraystretch{1.8}
\begin{tabular}{c|c}
\hline
spin-$S$ & $\mathsf{d}_{S}\mathsf{c}_{22S}|_{O(g^6)}$\\
\hline
2& $-768 + 112 \zeta_3 - 160 \zeta_5$\\
\hline
4 & $-\frac{442765625}{3500658} + \frac{386}{27}\zeta_3 - \frac{400 }{21}\zeta_5$\\
\hline
6 & $-\frac{1183056555847}{88944075000} + \frac{48286 }{37125}\zeta_3 - \frac{56}{33}\zeta_5$\\
\hline
8 & $-\frac{1270649655622342732745039}{1075922954067591630000000} + \frac{
 1039202363}{9932422500}\zeta_3 - \frac{6088}{45045}\zeta_5$\\
 \hline
 10 & $-\frac{7465848687069712820911408164847}{
  77747563297936585275804036000000} + \frac{
 8295615163}{1049947353000}\zeta_3 - \frac{2684}{264537}\zeta_5$\\
 \hline
\end{tabular}
\caption{3-loop result of the OPE coefficient from spin 2 to 10.}
\label{tab:dataDC}
\end{center}
\end{table}
Dividing the result by the known results of $\mathsf{c}_{22S}$ and taking the square, we find the OPE coefficient $(\mathsf{d}_S)^2$. The results up to $S=10$ are given below
\small{
\begin{align}
\mathsf{d}_2^2=&\,\frac{1}{3}-4 g^2+(56-24 \zeta_3)g^4 +16 (22 \zeta_3+5 \zeta_5-48) g^6\,,\\\nonumber
\mathsf{d}_4^2=&\,\frac{1}{35}-\frac{205}{441}g^2+ \left(\frac{70219}{9261}-\frac{20 \zeta_3}{7}\right)g^4+\left(\frac{58868 \zeta_3}{1323}+\frac{200 \zeta_5}{21}-\frac{200151970}{1750329}\right)g^6 \,,\\\nonumber
\mathsf{d}_6^2=&\,\frac{1}{462}-\frac{1106}{27225} g^2+\left(\frac{772465873}{1078110000}-\frac{14 \zeta_3}{55}\right)g^4 + \left(\frac{316477 \zeta_3}{81675}+\frac{28 \zeta_5}{33}-\frac{1001837354497}{88944075000}\right)g^6\,,\\\nonumber
\mathsf{d}_8^2=&\,\frac{1}{6435}-\frac{14380057}{4509004500} g^2+ \left(\frac{5048546158688587}{85305405235050000}-\frac{1522 \zeta_3}{75075}\right)g^4\\\nonumber
&+\left(\frac{209601639281 \zeta_3}{710168208750}+\frac{3044 \zeta_5}{45045}-\frac{256288425994027633489541}{268980738516897907500000}\right)g^6 \,,\\\nonumber
\mathsf{d}^2_{10}=&\,\frac{1}{92378}-\frac{3313402433}{13995964873800} g^2+ \left(\frac{141793274806850941159}{31100584702491617040000}-\frac{671 \zeta_3}{440895}\right)g^4\\\nonumber
&+ \left(\frac{11578996460944 \zeta_3}{551091116905875}+\frac{1342 \zeta_5}{264537}-\frac{5796660491433231307517026775347}{77747563297936585275804036000000}\right)g^6
\end{align}
}
\normalsize{}

\subsection{Harmonic sum and large spin limit}
In order to write down the result of $\mathsf{d}_S^2$ in a closed form for arbitrary spin $S$, we can rewrite the results in terms of nested harmonic sums. 
\paragraph{Harmonic Sum} The harmonic sums are defined as\footnote{In this subsection, we use $j$ instead of $S$ to denote spin in order to avoid confusion with the harmonic sums.}
\begin{equation}
	S_a(j) \equiv\left\{\begin{array}{ll}
		\sum_{n=1}^j \frac{1}{n^{|a|}} & a \geq 0 \\
		\sum_{n=1}^j \frac{(-1)^n}{n^{|a|}} & a<0
	\end{array}, \quad S_{a, b, \ldots}(j) \equiv \begin{cases}\sum_{n=1}^j \frac{1}{n^{|a|}} S_{b, \ldots}(n) & a \geq 0 \\
		\sum_{n=1}^j \frac{(-1)^n}{n^{|a|}} S_{b, \ldots}(n) & a<0\end{cases}\right.
\end{equation}
For simplicity, we will omit  the argument of the harmonic sums in what follows. The main idea is to write down an ansatz for the result as a linear combination of harmonic sums with certain weights. In order to fix the coefficients, we calculate the structure constant up to sufficiently high spin. For a given weight, we can choose a basis for all the harmonic sums. For example, the basis at weight 6 contains 486 independent harmonic sums. Such a basis can be found with the aide of certain packages like \texttt{HarmonicSum}\cite{Ablinger:2014rba}. Once all the coefficients are fixed, its correctness can be tested by comparing with even higher spin results. At each loop order, we make the uniform transcendentality ansatz with transcendental degree $2n$ at $n$-loops. Such a procedure has been carried out up to two-loop \cite{Jiang:2019xdz}. With our result, we can push this calculation to three loops, the result is given by
\begin{equation}
\label{eq:resSum}
\left(\frac{\mathsf{d}_{j}}{\left.\mathsf{d}_{ j}\right|_{\text{tree }}}\right)^2=\texttt{prefactor}\times\left[1-4 g^2 S_2+8 g^4\left(d_{2,4}+d_{2,1}\zeta_3\right)+64 g^6\left(d_{3,6}+d_{3,3}\zeta_3 +d_{3,1}\zeta_5\right)\right],
\end{equation}
with
\begin{align}
d_{2,1}=&\,-6S_1\,\\
d_{2,4}=&\,5 S_{-4}+8 S_{-3} S_1+4 S_{-2}\left(S_1\right)^2+2 S_{-2} S_2+2\left(S_2\right)^2+8 S_1 S_3 \\\nonumber
& +7 S_4-8 S_{-3,1}-8 S_1 S_{-2,1}-6 S_{-2,2}-4 S_{1,3}-4 S_{3,1}+8 S_{-2,1,1}\,
\end{align}
and
\begin{align}
d_{3,1}=&\,\frac{5}{2}S_1\,\\
d_{3,3}=&\,-2 S_{-2,1}-\frac{4 S_1^3}{3}+6 S_2 S_1+S_{-3}+\frac{4 S_3}{3},\\
d_{3,6}=&\frac{4}{3} S_1^3 S_{-2,1}+16 S_1^2 S_{-3,1}+10 S_1^2 S_{-2,2}-8 S_1^2 S_{-2,1,1}+31 S_1 S_{-4,1}+2 S_{-2} S_1 S_{-2,1}\\\nonumber
        &-12 S_2 S_1 S_{-2,1}-2 S_1 S_{-2,3}-4 S_1 S_{2,-3}+24 S_{-2} S_1 S_{2,1}-36 S_1 S_{-3,1,1}+4 S_1 S_{-2,1,-2}\\\nonumber
        &-24 S_1 S_{-2,2,1}-24 S_1 S_{2,1,-2}+24 S_1 S_{-2,1,1,1}+2 S_{-2,1}^2-8 S_{-5,1}+17 S_{-4,2}+2 S_{-2} S_{-3,1}\\\nonumber
        &-16 S_2 S_{-3,1}-3 S_{-3,3}-2 S_{-3} S_{-2,1}+\frac{38}{3} S_3 S_{-2,1}+2 S_{-2} S_{-2,2}+8 S_2 S_{-2,2}-6 S_{-2} S_{3,1}\\\nonumber
        &-21 S_{4,-2}-4 S_{-2} S_{-2,1,1}+16 S_2 S_{-2,1,1}-20 S_{-2,2,2}+6 S_{-2,3,1}+26 S_{2,-3,1}+6 S_{3,1,-2}\\\nonumber
        &+40 S_{-3,1,1,1}-28 S_{2,-2,1,1}-32 S_{-2,1,1,1,1}-\frac{14}{3} S_{-3} S_1^3-2 S_{-2}^2 S_1^2-17 S_{-4} S_1^2-6 S_{-2} S_2 S_1^2\\\nonumber
        &-3 S_4 S_1^2+\frac{7}{2} S_{-5} S_1-5 S_{-3} S_{-2} S_1+14 S_{-3} S_2 S_1-8 S_{-2} S_3 S_1-6 S_2 S_3 S_1-\frac{17 S_5 S_1}{2}\\\nonumber
        &-S_2^3+\frac{S_{-3}^2}{2}-2 S_{-2} S_2^2-S_3^2+4 S_{-6}-S_{-4} S_{-2}-\frac{1}{2} S_{-2}^2 S_2-\frac{13}{2} S_{-4} S_2-\frac{34}{3} S_{-3} S_3\\\nonumber
        &+18 S_{-2} S_4-3 S_2 S_4-\frac{5 S_6}{2}.
\end{align}

The prefactor is given by 
\begin{equation}
\texttt{prefactor}=\frac{\Gamma (2 j+1) \Gamma \left(j+\frac{\gamma }{2}+1\right)^2}{\Gamma (j+1)^2 \Gamma (2 j+\gamma +1)}-\frac{1}{4}\left(-\gamma^2+(S_{1}(2j)-S_{1}(j))\gamma^3 \right)\zeta_2-\frac{\gamma ^3 \zeta_3}{4},
\end{equation}
where $\gamma=\delta\Delta_{\mathcal{K}}$ is anomalous dimension.
\paragraph{Large Spin Limit} Another advantage of the harmonic sum representation is that it allows us to make analytic continuation and extract the large spin behavior of the structure constant. The details are delegated to appendix~\ref{app:LargeSpin}. Here we simply present the final answer at leading order:
\begin{equation}
 \begin{aligned}     \ln\left[\left(\frac{\mathsf{d}_{j}}{\left.\mathsf{d}_{j}\right|_{\text {tree }}}\right)^2\right]=&-4 g^2 \Big(\zeta _2+2 \ln{2} \ln{S}\Big)+\frac{8}{5} g^4 \Big(8 \zeta _2^2+15 \zeta _3 \ln{2}+10 \zeta _2 \ln{2} \ln{S}-15 \zeta _3 \ln{S}\Big)\\
        &+\frac{16}{105} g^6 \Big(-449 \zeta _2^3+385 \zeta _3^2-210 \zeta _2 \zeta _3 \ln{2}-1050 \zeta _5 \ln{2}\\
        &-462 \zeta _2^2 \ln{2} \ln{S}+420 \zeta _2 \zeta _3 \ln{S}+1050 \zeta _5 \ln{S}\\
        &-252 \zeta _2^2 (\ln{S})^2 +840 \zeta _2 \ln{2}(\ln{S})^3-280 \zeta _3 (\ln{S})^3\Big),
  \end{aligned}
\end{equation}
where $\ln{S}=\ln{j}+\gamma_E$ and $\gamma_{E}$ is the Euler constant. Up to two-loop, the large spin limit exhibit only $\ln S$ behavior. However, as we can see from the result, at three-loop order we start to have contributions like $(\ln S)^2$ and $(\ln S)^3$.

\subsection{Discussions}
Intriguingly, we find that up to three-loop order, our results of $\mathsf{d}_S\mathsf{c}_{22S}$ in Table~\ref{tab:dataDC} coincide exactly with $\mathsf{c}_{44S}^2$ (see \emph{e.g.} Table-1 of \cite{Basso:2015eqa}). 
\begin{align}
\label{eq:dsCC}
\mathsf{d}_{S}\mathsf{c}_{22S}=\mathsf{c}_{44S}^2
\end{align}
At three-loop order, we actually have
\begin{align}
\mathsf{c}_{44S}^2=\mathsf{c}^2_{LLS},\qquad L\ge 4\,.
\end{align}
This is clear from the point of view of hexagon form factors. The OPE coefficients of $\mathsf{c}_{LLS}$ have the same asymptotic contribution and the adjacent wrapping contributions. What makes the difference is the bottom or opposite wrapping corrections. For $L\ge 4$, bottom wrappings do not contribute at three-loop order. However, for $L=2$ we do have non-zero three-loop contribution and we have
\begin{align}
\mathsf{c}_{44S}^2=\mathsf{c}_{22S}^2-\texttt{bottom wrapping}
\end{align}
The relation \eqref{eq:dsCC} shows that there is an intimate relation between the OPE coeffcients of single trace operators and giant gravitons with twist-two operators. It is tempting to conjecture that at higher loop orders, similar relation $\mathsf{d}_S\mathsf{c}_{22S}=\mathsf{c}_{LLS}^2$ holds for $L>4$. If this were the case, it implies that there should be an intimate connection between the worldsheet $g$-function approach and the hexagon form factor approach to the OPE coefficients. One can expect to use the $g$-function to partially resum the hexagon mirror corrections. This would be fascinating to check.

\section{Conclusions and Outlook}
\label{sec:conclude}
In this paper, we computed the four-point function with two maximal giant gravitons and two length-2 single-trace chiral BPS operators up to three-loop order in the planar limit. Our main result is \eqref{eq:finalRes} written in terms of known three-loop conformal integrals. From the four-point function, we extract the OPE coefficients of two giant graviton and twist-two operators with arbitrary spin. The result is given in terms of harmonic sums \eqref{eq:resSum}, from which we can extract the large spin behavior of the OPE coefficient. We find the contributions of the form $(\ln S)^2$ and $(\ln S)^3$ start to contribute at three-loop order.\par

From the explicit results, we find an intriguingly simple relation between the OPE coefficient of giant gravitons and single-trace operators, given in \eqref{eq:dsCC}. This relation hints a deep connection between these two types of OPE coefficients. From integrability point of view, these two kinds of OPE coefficients are computed by different approaches. The giant graviton OPE coefficient is computed by the worldsheet $g$-function approach. The advantage of this approach is that all the finite size corrections can be taken into account, thanks to thermodynamic Bethe ansatz. On the other hand, the OPE coefficient of single-trace operators are computed by hexagon form factors. A systematic approach to take into account all finite size corrections is still missing at the moment, although important partial progress has been made (see for example \cite{Jiang:2016ulr,Kostov:2019stn,Belitsky:2020qrm,Bargheer:2019kxb,Bargheer:2019exp}). In particular, there is an all-loop conjecture for the OPE coefficients $\mathsf{c}_{22S}$ \cite{Basso:2022nny}. It would be interesting to make a more explicit connection between these two approaches.\par

There are many future questions that need to be addressed. The first thing would be test the $g$-function prediction from integrability using our field theoretical result. In particular, we will see whether some kind of wrapping corrections would show up at this loop order, or asymptotic result is sufficient like in the spectral problem. This work has been initiated and we would like to report it elsewhere.\par

Another important direction is to extent the field theoretical result to include BPS operators with larger length, namely compute the four-point functions of $\langle\mathcal{D}(x_1)\mathcal{D}(x_2)\mathcal{O}_j(x_3)\mathcal{O}_k(x_4)\rangle$ for $j,k>2$ up to three-loop order. At the same time, it is also interesting to consider four-point functions of giant gravitons, namely $\langle\mathcal{D}(x_1)\ldots\mathcal{D}(x_4)\rangle$. Such correlation functions is only known up to one-loop, but we believe it should be possible to push it to at least three-loop orders.\par

Finally a more challenging but obvious next step is to compute all the aforementioned four-point functions to 4- and 5-loops, catching up with the results of the single-trace operators. However, the unknown coefficients grows rapidly. Our current strategy which uses lower loop data can only fix part of the unknown coefficients. To fix the full result, we shall need more constraints from other principles.

\section*{Acknowledgements}
We would like to thank Shota Komatsu and Xinan Zhou for very helpful discussions. We also thank Bukhard Eden, Claude Duhr for the support on the computation of conformal integrals. Furthermore, we are grateful to Yingxuan Xu because of his help on the numeric checks of conformal integrals. The work of YJ is partly supported by Startup Funding no. 3207022217A1 of Southeast University. YZ is supported from the NSF of China through Grant No. 11947301, 12047502, 12075234 and 12247103.

\appendix
\section{Perturbative results up to two-loop}
\label{app:pert2loop}
In this section, we review the results for the giant graviton two-point function up to two-loop order. We will compare the results in the $s$- and $t$-channels by direct computation and OPE analysis.

\paragraph{$s$-channel} In the $s$-channel direct computation, for the result written in terms an ansatz, we have
\begin{align}
\lim_{z,\bar{z}\to 0}\frac{G_{\{2,2\}}^{(\ell)}}{\tilde{R}_{1234}(d_{12})^{N-2}}=\lim_{u\to0\atop v\to 1}\frac{G_{\{2,2\}}^{(\ell)}}{\tilde{R}_{1234}(d_{12})^{N-2}}=\sum_{k} c_k\,\lim_{u\to 0\atop v\to 1}I^{(l)}_k
\end{align}
where we take the proper limit for the conformal integrals at $\ell$-loop order. On the other hand, from OPE analysis, we have
\begin{align}
\lim_{x_1\to x_2\atop x_3\to x_4}\frac{G_{\{2,2\}}^{(\ell)}}{\tilde{R}_{1234}(d_{12})^{N-2}}=\left.\mathsf{d}_{\mathcal{K}}(g)\mathsf{c}_{\mathcal{K}}(g) u^{\delta\Delta/2}\right|_{O(g^{2\ell})}\,.
\end{align}
We make the following expansion
\begin{align}
\delta\Delta(g)=\sum_{n=1}^{\infty}\gamma_n\,g^{2n},\qquad \mathsf{d}_{\mathcal{K}}(g)\mathsf{c}_{\mathcal{K}}(g)=\sum_{n=0}^{\infty}C_n\,g^{2n}
\end{align}
From tree level computation, we find that $C_0=1/3$. From integrability, $\gamma_1=12$.

\paragraph{$t$-channel} In the $t$-channel, from the ansatz we have
\begin{align}
\lim_{z,\bar{z}\to 1}\frac{G_{\{2,2\}}^{(\ell)}}{\tilde{R}_{1234}(d_{12})^{N-2}}=\lim_{u\to1\atop v\to 0}\frac{G_{\{2,2\}}^{(\ell)}}{\tilde{R}_{1234}(d_{12})^{N-2}}=\sum_{k} c_k\,\lim_{u\to 1\atop v\to 0}I^{(l)}_k
\end{align}
where again we need to take the proper limit of the conformal integral. From OPE analysis, we have 
\begin{align}
\lim_{x_1\to x_4\atop x_3\to x_2}\frac{G_{\{2,2\}}^{(\ell)}}{\tilde{R}_{1234}(d_{12})^{N-2}}=\left.(\mathsf{c}_{\mathcal{O}_{\text{open}}}(g))^2\,v^{\delta\Delta_{\text{open}}/2}\right|_{O(g^{2\ell})}\,.
\end{align}
We make the following expansion
\begin{align}
\delta\Delta_{\text{open}}=\sum_{n=1}^{\infty}\tilde{\gamma}_n\,g^{2n},\qquad \left(\mathsf{c}_{\text{Open}}(g)\right)^2=\sum_{n=0}^{\infty}\widetilde{C}_ng^{2n}\,.
\end{align}
From tree level computation, we find that $\widetilde{C}_0=1$\,.

\subsection{One-loop}
At one loop, the ansatz \eqref{eq:constructGP} is fixed up to a constant. Let us denote $P^{(1)}=c/(-4\pi^2)$. Then \eqref{eq:constructGP} becomes
\begin{align}
G_{\{2,2\}}^{(1)}=R_{1234}(d_{12})^{N-2}\times\frac{1}{(-4\pi^2)}\int\frac{c\,\rd^4 x_5}{x_{15}^2x_{25}^2x_{35}^2x_{45}^2}\,.
\end{align}
The integral can be computed analytically and is given by the one-loop conformal integral
\begin{align}
F^{(1)}(z,\bar{z})=\frac{x_{13}^2x_{24}^2}{\pi^2}\int\frac{\rd^4 x_5}{x_{15}^2x_{25}^2x_{35}^2x_{45}^2}
=\frac{1}{z-\bar{z}}\left(2\text{Li}_2(z)-2\text{Li}_2(\bar{z})+\ln(z\bar{z})\ln\frac{1-z}{1-\bar{z}}\right)
\end{align}

\paragraph{$s$-channel} From direct computation, in the $s$-channel we have
\begin{align}
\label{eq:oneloopLog1}
\lim_{z,\bar{z}\to 0}F^{(1)}(z,\bar{z})=-\ln(z\bar{z})+2+\cdots=-\ln u+2+\cdots
\end{align}
Therefore we have
\begin{align}
\lim_{z,\bar{z}\to 0}\frac{G_{\{2,2\}}^{(1)}}{\tilde{R}_{1234}(d_{12})^{N-2}}=\frac{c}{4}\ln u-\frac{c}{2}\,.
\end{align}
On the other hand, from direct OPE analysis, we have
\begin{align}
\label{eq:oneloopLog2}
\lim_{z,\bar{z}\to 0}\frac{G_{\{2,2\}}^{(1)}}{\tilde{R}_{1234}(d_{12})^{N-2}}=\frac{1}{8}C_0\gamma_1\,\ln u+\frac{C_1}{4}\,.
\end{align}
From tree level computation and integrability, we know that $C_0=1/3$ and $\gamma_1=12$. Comparing the coefficients of $\log u$ of \eqref{eq:oneloopLog1} and \eqref{eq:oneloopLog2}, we fix the constant $c=2$. Plugging back, we find that this gives us the one-loop result $C_1=-2c=-4$. Therefore we got
\begin{align}
\mathsf{d}_{\mathcal{K}}(g)\mathsf{c}_{\mathcal{K}}(g)=\frac{1}{3}-4g^2\,.
\end{align}

\paragraph{$t$-channel} We can now expand the result in the $t$-channel, which gives us $\tilde{\gamma}_1$ and $\widetilde{C}_1$. By direct expansion, we find that
\begin{align}
\label{eq:oneloopLog3}
\lim_{z,\bar{z}\to 1}\frac{G_{\{2,2\}}^{(1)}}{\tilde{R}_{1234}(d_{12})^{N-2}}=\frac{1}{2}\ln v-1\,.
\end{align}
On the other hand, from OPE analysis, we find that
\begin{align}
\label{eq:oneloopLog4}
\lim_{z,\bar{z}\to 1}\frac{G_{\{2,2\}}^{(1)}}{\tilde{R}_{1234}(d_{12})^{N-2}}=\frac{1}{8}\widetilde{C}_0\tilde{\gamma}_1\,\ln v+\frac{\widetilde{C}_1}{4}\,.
\end{align}
Comparing \eqref{eq:oneloopLog3} and \eqref{eq:oneloopLog4} and using $\widetilde{C}_0=1$, we obtain
\begin{align}
\tilde{\gamma}_1=4,\qquad \widetilde{C}_1=-4\,.
\end{align}
Therefore at one-loop order, we find that
\begin{align}
\delta\Delta_{\text{open}}=4g^2+\ldots,\qquad \left(\mathsf{c}_{\text{Open}}(g)\right)^2=1-4g^2+\ldots
\end{align}

\subsection{Two-loop}
\begin{figure}[h!]
\centering
\includegraphics[scale=0.4]{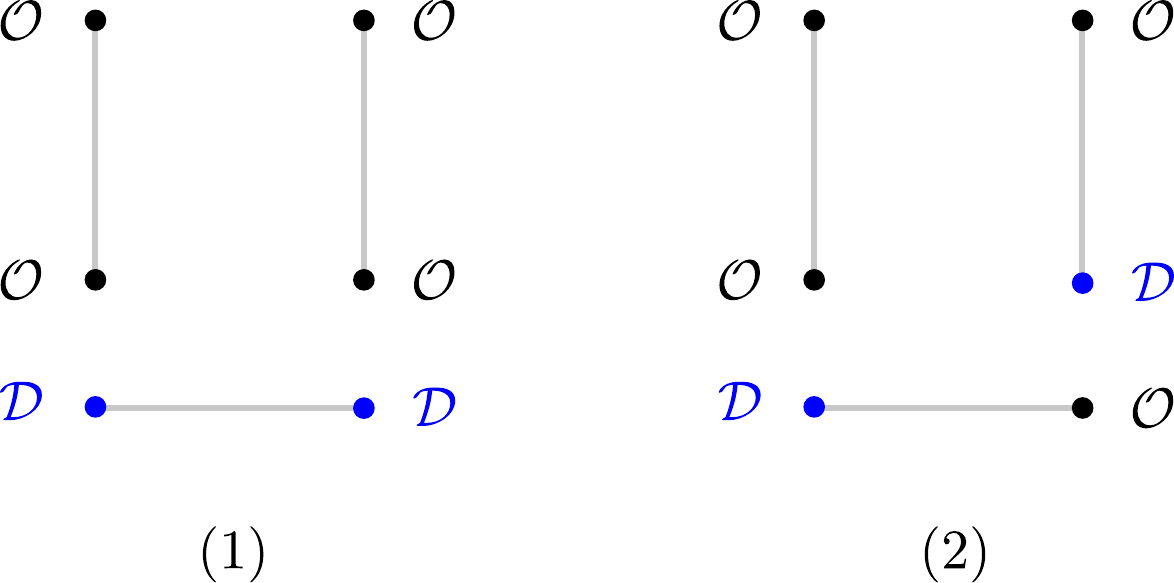}
\caption{Two diagrams for the two-loop ansatz.}
\label{fig:twoLoop}
\end{figure}

At two-loop order, we need to draw all different diagrams with 6 bullet and each bullet is connected to another one. Up to $S_2\times S_4$ permutations, we have two types of diagrams in figure~\ref{fig:twoLoop}.
These two diagrams correspond to two polynomials
\begin{align}
&P_1^{(2)}(x_i)=x_{12}^2x_{34}^2x_{56}^2+S_2\times S_4\text{ permutations}\,,\\\nonumber
&P_2^{(2)}(x_i)=x_{13}^2x_{24}^2x_{56}^2+S_2\times S_4\text{ permutations}\,,
\end{align}
where $S_2$ permutation permutes $x_1,x_2$ and $S_4$ permutations permutes $x_3,x_4,x_5,x_6$. Written down more explicitly, we have
\begin{align}
P_1^{(2)}(x_i)=16\,p_1(x_i),\qquad P_2^{(2)}(x_i)=4\,p_2(x_i)
\end{align}
where
\begin{align}
p_1(x_i)=&\,x_{12}^2(x_{34}^2x_{56}^2+x_{35}^2x_{46}^2+x_{36}^2x_{45}^2)\,,\\\nonumber
p_2(x_i)=&\,x_{16}^2x_{25}^2x_{34}^2+x_{15}^2x_{26}^2x_{34}^2+x_{16}^2x_{24}^2x_{35}^2+x_{15}^2x_{24}^2x_{36}^2\\\nonumber
&\,+x_{14}^2x_{26}^2x_{35}^2+x_{14}^2x_{25}^2x_{36}^2+x_{16}^2x_{23}^2x_{45}^2+x_{15}^2x_{23}^2x_{46}^2\\\nonumber
&\,+x_{13}^2x_{26}^2x_{45}^2+x_{13}^2x_{25}^2x_{46}^2+x_{14}^2x_{23}^2x_{56}^2+x_{13}^2x_{24}^2x_{56}^2\,.
\end{align}
Therefore we make the ansatz
\begin{align}
P^{(2)}=\frac{c_1P^{(2)}_1+c_2P_2^{(2)}}{(-4\pi^2)^2}=\frac{c_1\,p_1+c_2\,p_2/4}{\pi^4}\,.
\end{align}
The ansatz has two unknown coefficients which we denote by $c_1$ and $c_2$\footnote{Note that our convention here is slightly different from the ones in \cite{Jiang:2019zig,Jiang:2019xdz} The coefficient $c_2$ is different by a factor of 4.}. The corresponding two-loop integral can be written in terms of conformal 1- and 2-loop integrals as follows
\begin{align}
\frac{G_{\{2,2\}}^{(2)}}{\tilde{R}_{1234}(d_{12})^{N-2}}=&\,\left(c_1\,z\bar{z}+\frac{c_2}{4}(1-z)(1-\bar{z})+\frac{c_2}{4}\right)\left(F^{(1)}(z,\bar{z})\right)^2\\\nonumber
&\,+c_2\,F_z^{(2)}+2(c_1+c_2/4)F_{1-z}^{(2)}+c_2\,F^{(2)}_{\frac{z}{z-1}}\,,
\end{align}
where $F^{(1)}$ is the one-loop conformal integral. The various two-loop conformal integrals are defined by
\begin{align}
F_z^{(2)}(z,\bar{z})=F^{(2)}(z,\bar{z}),\qquad F_{1-z}^{(2)}=F^{(2)}(1-z,1-\bar{z}),\qquad F^{(2)}_{\frac{z}{z-1}}=\frac{F^{(2)}\left(\frac{z}{z-1},\frac{\bar{z}}{\bar{z}-1}\right)}{(1-z)(1-\bar{z})}\,,
\end{align}
where $F^{(2)}$ is the two-loop conformal integral defined by
\begin{align}
F^{(2)}(z,\bar{z})=&\,\frac{x_{13}^2x_{24}^2x_{14}^2}{\pi^4}\int\frac{\rd^4 x_5\rd^4 x_6}{x_{15}^2x_{25}^2x_{45}^2x_{56}^2x_{16}^2x_{36}^2x_{46}^2}\\\nonumber
=&\,\frac{1}{z-\bar{z}}\left[\frac{\ln(z\bar{z})^2}{2}\left(\text{Li}_2(z)-\text{Li}_2(\bar{z})\right)
-3\log(z\bar{z})\left(\text{Li}_3(z)-\text{Li}_3(\bar{z})\right)+6\left(\text{Li}_4(z)-\text{Li}_4(\bar{z})\right)\right]
\end{align}

\paragraph{$s$-channel OPE} In the $s$-channel, we have
\begin{align}
\lim_{z,\bar{z}\to 0}F_z^{(2)}
=\lim_{z,\bar{z}\to 0}F_{\frac{z}{z-1}}^{(2)}=\frac{1}{2}\ln^2 u-3\ln u+6\,,\qquad\lim_{z,\bar{z}\to 0}F^{(2)}_{1-z}=6\zeta_3\,.
\end{align}
Using these results, the $s$-channel limit gives
\begin{align}
\label{eq:ansatz2loop}
\lim_{x_1\to x_2\atop x_3\to x_4}\frac{G_{\{2,2\}}^{(2)}}{\tilde{R}_{1234}(d_{12})^{N-2}}=\frac{3}{2}c_2(\ln u)^2-8c_2\,\ln u+12(c_1+c_2/4)\zeta(3)+14c_2\,.
\end{align}
On the other hand, we have
\begin{align}
\label{eq:expanddcu2loop}
\left.\mathsf{d}_{\mathcal{K}}(g)\mathsf{c}_{\mathcal{K}}(g){u^{\delta \Delta/2}}\right|_{O(g^4)}=&\,\frac{1}{8}C_0\gamma_1^2\,(\ln u)^2+\frac{1}{2}(C_1\gamma_1+C_0\gamma_2)(\ln u)+C_2
\end{align}
Using the results from the previous section and the known anomalous dimension of the Konishi operator, we have
\begin{align}
C_0=\frac{1}{3},\qquad C_1=-4,\qquad \gamma_1=12,\qquad \gamma_2=-48\,.
\end{align}
Comparing \eqref{eq:ansatz2loop} and \eqref{eq:expanddcu2loop}, we find the equations
\begin{align}
\frac{3}{2}c_2=\frac{1}{8}C_0\gamma_1^2=6,\qquad 
-8c_2=\frac{1}{2}(C_1\gamma_1+C_0\gamma_2)
\end{align}
These are two equations for $c_2$. They are compatible and gives $c_2=4$. To determine the other unknown coefficient $c_1$, we consider the $t$-channel limit.

\paragraph{$t$-channel OPE} In the $t$-channel limit, we have
\begin{align}
\lim_{z,\bar{z}\to 1}F_z^{(2)}=6\zeta_3\,,\qquad \lim_{z,\bar{z}\to 1}F^{(2)}_{1-z}=\lim_{z,\bar{z}\to 1}F^{(2)}_{\frac{z}{z-1}}=\frac{1}{2}(\ln v)^2-3\ln v+6
\end{align}
Using these results, in the $t$-channel limit, we obtain
\begin{align}
\label{eq:tchannel2loop}
\lim_{z,\bar{z}\to 1}\frac{G_{\{2,2\}}^{(2)}}{\tilde{R}_{1234}(d_{12})^{N-2}}=(2c_1+c_2)(\ln v)^2-\left(10c_1+\frac{11c_2}{2} \right)\ln v+6c_2\zeta_3+(16c_1+10c_2)\,.
\end{align}
From the OPE limit, we obtain
\begin{align}
\label{eq:twoloopLog4}
\lim_{z,\bar{z}\to 1}\frac{G_{\{2,2\}}^{(2)}}{\tilde{R}_{1234}(d_{12})^{N-2}}=\frac{1}{8}\widetilde{C}_0\tilde{\gamma}_1^2\,(\ln v)^2+\frac{1}{2}\left(\widetilde{C}_1\tilde{\gamma}_1+\widetilde{C}_0\tilde{\gamma}_2\right)\ln v+\widetilde{C}_2\,.
\end{align}
From the one-loop result, we find that
\begin{align}
\widetilde{C}_0=1,\qquad \widetilde{C}_1=-4,\qquad \tilde{\gamma}_1=4\,.
\end{align}
Comparing the coefficients of $(\ln v)^2$, we find that
\begin{align}
2c_1+c_2=\frac{1}{8}\widetilde{C}_0\tilde{\gamma}_1^2\quad\Longrightarrow\quad 2c_1+4=2
\end{align}
which leads to $c_1=-1$. Plugging $c_1=-1,c_2=4$ into \eqref{eq:tchannel2loop}, we find that
\begin{align}
\tilde{\gamma}_2=-8,\qquad \widetilde{C}_2=24+24\zeta_3\,.
\end{align}
Plugging into \eqref{eq:ansatz2loop}, we further find that
\begin{align}
C_2=56\,.
\end{align}
To sum up, up to two-loop order, we find that
\begin{align}
P^{(2)}(x_i)=\frac{p_2(x_i)-p_1(x_i)}{\pi^4}\,.
\end{align}
In the $s$-channel, we have
\begin{align}
\gamma(g)=12g^2-48g^4+\ldots\,,\qquad \mathsf{d}_{\mathcal{K}}(g)\mathsf{c}_{\mathcal{K}}(g)=\frac{1}{3}-4g^2+56g^4+\ldots.
\end{align}
In the $t$-channel we have
\begin{align}
\delta\Delta_{\text{open}}=&\,4g^2-8g^4+\ldots,\\\nonumber
\left(\mathsf{c}_{\text{Open}}(g)\right)^2=&\,1-4g^2+24(1+\zeta_3)g^4+\ldots
\end{align}

\section{Three-loop integrals}
\label{app:integrals}
In this appendix, we list the three-loop integrals that are needed in the main text. These integrals were defined in [...]
\begin{align}
\label{eq:3loopInt}
T(1,2;3,4)=&\,\frac{x_{34}^2}{(-4\pi^2)^3}\int\frac{\rd^4 x_5\rd^4 x_6\rd^4 x_7\,x_{17}^2}
{(x_{15}^2x_{35}^2)(x_{16}^2x_{46}^2)(x_{37}^2x_{27}^2x_{47}^2)x_{56}^2x_{57}^2x_{67}^2}\,,\\\nonumber
E(1,2;3,4)=&\,\frac{x_{23}^2x_{24}^2}{(-4\pi^2)^3}\int\frac{\rd^4 x_5\rd^4 x_6\rd^4 x_7\,x_{16}^2}
{(x_{15}^2x_{25}^2x_{35}^2)x_{56}^2(x_{26}^2x_{36}^2x_{46}^2)x_{67}^2(x_{17}^2x_{27}^2x_{47}^2)}\,,\\\nonumber
L(1,2;3,4)=&\,\frac{x_{34}^4}{(-4\pi^2)^3}\int\frac{\rd^4 x_5\rd^4 x_6\rd^4 x_7}{(x_{15}^2x_{35}^2x_{45}^2)x_{56}^2(x_{36}^2x_{46}^2)x_{67}^2(x_{27}^2x_{37}^2x_{47}^2)}\,,\\\nonumber
gh(1,2;3,4)=&\,\frac{x_{12}^2x_{34}^4}{(-4\pi^2)^3}\int\frac{\rd^4 x_5\rd^4 x_6\rd^4 x_7}
{(x_{15}^2x_{25}^2x_{35}^2x_{45}^2)(x_{16}^2x_{36}^2x_{46}^2)(x_{27}^2x_{37}^2x_{47}^2)x_{67}^2}\,,\\\nonumber
H(1,2;3,4)=&\,\frac{x_{14}^2x_{23}^2x_{34}^2}{(-4\pi^2)^3}\int\frac{\rd^4 x_5\rd^4 x_6\rd^4 x_7\,x_{57}^2}
{(x_{15}^2x_{25}^2x_{35}^2x_{45}^2)x_{56}^2(x_{36}^2x_{46}^2)x_{67}^2(x_{17}^2x_{27}^2x_{37}^2x_{47}^2)}\,.
\end{align}
Notice that $gh(1,2;3,4)$ indeed factorizes into the product of two integrals, as the notation indicates
\begin{align}
gh(1,2;3,4)=x_{12}^2x_{34}^2\,g(1,2,3,4)\times h(1,2;3,4)
\end{align}
where
\begin{align}
g(1,2,3,4)=&\,-\frac{1}{4\pi^2}\int\frac{\rd^4 x_5}{x_{15}^2x_{25}^2x_{35}^2x_{45}^2}\,,\\\nonumber
h(1,2;3,4)=&\,\frac{x_{34}^2}{(4\pi^2)^2}\int\frac{\rd^4 x_5\rd^4 x_6}{(x_{15}^2x_{35}^2x_{45}^2)x_{56}^2(x_{26}^2x_{36}^2x_{46}^2)}\,.
\end{align}

\subsection{Symmetry of integrals}
The integrals defined in \eqref{eq:3loopInt} satisfy certain relations under re-ordering the labels. We list them here.
\paragraph{Manifest invariance} These are the symmetries that are already manifest at the level of integrands.
\begin{align}
&L(1,2;3,4)=L(2,1;3,4),\quad L(1,2;4,3)=L(1,2;3,4)\,,\\\nonumber
&H(1,2;3,4)=H(2,1;4,3),\quad E(1,2;4,3)=E(1,2;3,4)\,.
\end{align}
\paragraph{Flip identities}
\begin{align}
&L(1,2;3,4)=L(3,4;1,2),\quad E(1,2;3,4)=E(3,4;1,2),\\\nonumber 
&H(1,2;3,4)=H(3,4;1,2),\quad E(1,2;3,4)=E(2,1;3,4)\,.
\end{align}

\paragraph{Identity between integrals} Finally we have
\begin{align}
T(1,2;3,4)=L(1,2;3,4)\,.
\end{align}

\paragraph{Additional relations} There are some additional relations which are also useful in the main text.
\begin{align}
H(1,2;4,3)=\frac{1}{v}H(1,2;3,4)
\end{align}

\subsection{$s$-channel asymptotics}
In the main text, we need to use the value of the integrals in the $s$-channel OPE limit $x_1\to x_2$, $x_3\to x_4$, which is equivalent to $u\to 0$, $v\to 1$ in terms of conformal cross ratios. The leading order contribution in this limit has been worked out for the various integrals, which we list here. We first have
\begin{align}
x_{24}^4\,g(1,2,3,4)_s=&\,\frac{1}{4}\ln u-\frac{1}{2}\,,\\\nonumber
x_{24}^4\,h(1,2;3,4)_s=&\,\frac{3}{8}\zeta(3)\,,\\\nonumber
x_{24}^4\,h(1,3;2,4)_s=&\,\frac{1}{32}(\ln u)^2-\frac{3}{16}\ln u+\frac{3}{8}\,,\\\nonumber
x_{24}^4\,h(1,4;2,3)_s=&\,\frac{1}{32}(\ln u)^2-\frac{3}{16}\ln u+\frac{3}{8}\,.
\end{align}
which leads to
\begin{align}
x_{24}^4\,gh(1,2;3,4)_s=&\,\frac{3}{32}\zeta(3)\,\ln u-\frac{3}{16}\zeta(3)\,,\\\nonumber
x_{24}^4\,gh(1,3;2,4)_s=&\,\frac{1}{128}(\ln u)^3-\frac{1}{16}(\ln u)^2+\frac{3}{16}\ln u-\frac{3}{16}\,,\\\nonumber
x_{24}^4\,gh(1,4;2,3)_s=&\,\frac{1}{128}(\ln u)^3-\frac{1}{16}(\ln u)^2+\frac{3}{16}\ln u-\frac{3}{16}\,.
\end{align}
The rest of the asymptotic integrals are given by
\begin{align}
x_{24}^4\,L(1,2;3,4)_s=&\,-\frac{5}{16}\zeta(5)\,,\\\nonumber
x_{24}^4\,L(1,3;2,4)_s=&\,\frac{1}{384}(\ln u)^3-\frac{1}{32}(\ln u)^2+\frac{5}{32}\ln u-\frac{5}{16}\,,\\\nonumber
x_{24}^4\,L(1,4;2,3)_s=&\,\frac{1}{384}(\ln u)^3-\frac{1}{32}(\ln u)^2+\frac{5}{32}\ln u-\frac{5}{16}\,.
\end{align}
\begin{align}
x_{24}^4\,E(1,2;3,4)_s=&\,\frac{1}{192}(\ln u)^3-\frac{3}{64}(\ln u)^2+\frac{5}{32}\ln u-\frac{5}{16}\zeta(5)+\frac{3}{32}\zeta(3)-\frac{5}{32}\,,\\\nonumber
x_{24}^4\,E(1,3;2,4)_s=&\,\frac{3}{32}\zeta(3)\,\ln u-\frac{3}{16}\zeta(3)\,,\\\nonumber
x_{24}^4\,E(1,4;2,3)_s=&\,\frac{3}{32}\zeta(3)\,\ln u-\frac{3}{16}\zeta(3)\,.
\end{align}
\begin{align}
x_{24}^4\,H(1,2;3,4)_s=&\,\frac{3}{16}\zeta(3)\,\ln u-\frac{3}{16}\zeta(3)\,,\\\nonumber
x_{24}^4\,H(1,3;2,4)_s=&\,\frac{1}{192}(\ln u)^3-\frac{1}{16}(\ln u)^2+\frac{9}{32}\ln u+\frac{1}{4}\zeta(3)-\frac{7}{16}\,,\\\nonumber
x_{24}^4\,H(1,4;2,3)_s=&\,\frac{1}{192}(\ln u)^3-\frac{1}{16}(\ln u)^2+\frac{9}{32}\ln u+\frac{1}{4}\zeta(3)-\frac{7}{16}\,,
\end{align}
where we use an index `$s$' to denote the $s$-channel OPE limit. The asymptotic results above can be examined by numerical calculation via {\sc AMFlow} and {\sc Fiesta} \cite{Liu:2022chg,Smirnov:2021rhf}.

\subsection{$t$-channel asymptotics}
The asymptotic behavior of the three-loop integral in $t$-channel can be obtained by the following trick. The $t$-channel limit $x_1\to x_4$, $x_2\to x_3$ can be obtained by: (1) Swap $x_2$ and $x_4$; (2) Take the $s$-channel OPE limit in the new configuration. Notice that when swapping $x_2$ and $x_4$, we also swap the role of $u$ and $v$. Since all integrals involved are single-valued harmonic polylogarithms, this trick allows us to make use of the $s$-channel OPE which was derived in the previous subsection. For example, the $t$-channel asymptotic behavior of $H(1,2;3,4)$ can be obtained as
\begin{align}
H(1,2;3,4)_t\overset{x_2\leftrightarrow x_4}{=}\left.H(1,4;3,2)_s\right|_{u\to v}=\left.u H(1,4;2,3)_s\right|_{u\to v}\,.
\end{align}
Therefore, the asymptotic behavior of $H(1,2;3,4)$ in the $t$-channel is
\begin{align}
x_{24}^4\,H(1,2;3,4)_t= v \left(\frac{1}{192} \ln(v)^3 - \frac{1}{16} \ln(v)^2 + \frac{9}{32} \ln(v) + \frac{1}{4} \zeta(3)\right).
\end{align}
Using this trick, we obtain the $t$-channel asymptotics  of the three-loop integrals, which is listed as follows 
\begin{align}
x_{24}^4\,gh(1, 4; 2, 3)_t =&\,-\frac{3v}{16} \zeta(3) + \frac{3v}{32} \zeta(3) \ln v\,,\\\nonumber
x_{24}^4\,gh(1, 2; 3, 4)_t =&\,\frac{1}{128}(\ln v)^3-\frac{1}{16}(\ln v)^2+\frac{3}{16}\ln v-\frac{3}{16}\,,\\\nonumber
x_{24}^4\,gh(1, 3; 2, 4)_t =&\,\frac{1}{128}(\ln v)^3-\frac{1}{16}(\ln v)^2+\frac{3}{16}\ln v-\frac{3}{16}\,.
\end{align}
\begin{align}
x_{24}^4\,L(1, 4; 2, 3)_t=&\, -\frac{5}{16} \zeta(5)\,,\\\nonumber
x_{24}^4\,L(1, 3; 2, 4)_t=&\, \frac{1}{384} (\ln v)^3-\frac{1}{32}(\ln v)^2+ \frac{5}{32} \ln v-\frac{5}{16}\,,\\\nonumber
x_{24}^4\,L(1, 2; 3, 4)_t=&\, \frac{1}{384} (\ln v)^3-\frac{1}{32}(\ln v)^2+ \frac{5}{32} \ln v-\frac{5}{16}.
\end{align}
\begin{align}
x_{24}^4\,E(1, 4; 2, 3)_t=&\,\frac{1}{192}(\ln v)^3-\frac{3}{64}(\ln v)^2+\frac{5}{32}\ln v-\frac{5}{16}\zeta(5)+\frac{3}{32}\zeta(3)-\frac{5}{32} \\\nonumber
x_{24}^4\,E(1, 2; 3, 4)_t=&\,\frac{3}{32} \zeta(3)\ln v-\frac{3}{16}\zeta(3)\,, \\\nonumber
x_{24}^4\,E(1, 3; 2, 4)_t=&\,\frac{3}{32} \zeta(3)\ln v -\frac{3}{16}\zeta(3)\,,
\end{align}
\begin{align}
x_{24}^4\,H(1, 4; 2, 3)_t=&\,\frac{3}{16} \zeta(3)\ln v -\frac{3}{16}\zeta(3)\,,\\\nonumber
x_{24}^4\,H(1, 2; 3, 4)_t=&\,\frac{v}{192}(\ln v)^3-\frac{v}{16}(\ln v)^2+\frac{9v}{32}\ln v+ \frac{v}{4}\zeta(3)-\frac{7v}{16}\,,\\\nonumber
x_{24}^4\,H(1, 3; 2, 4)_t=&\,\frac{v}{192}(\ln v)^3-\frac{v}{16}(\ln v)^2+\frac{9v}{32}\ln v+ \frac{v}{4}\zeta(3)-\frac{7v}{16}\,,
\end{align}

\section{Large spin limit}
\label{app:LargeSpin}
The asymptotic behavior of harmonic sums can be computed by the method in \cite{Ablinger:2012ufz}. The main idea of this method is to use the relation between harmonic sums and Mellin transformation of harmonic polylogarithms. Mellin transformation is defined as
\begin{equation}
    \mathrm{M}(f(x),n)=\int^1_0 x^n f(x).
\end{equation}
Harmonic polylogarithms under this transformation have the following properties\cite{Ablinger:2012ufz}:
\begin{equation}
    \mathrm{M}\left(\frac{\mathrm{H}_{\boldsymbol{m}}(x)}{1-x},n+1\right)=-(n+1)\mathrm{M}(\mathrm{H}_{1,\boldsymbol{m}}(x),n),
\end{equation}
\begin{equation}
    \mathrm{M}\left(\frac{\mathrm{H}_{\textbf{m}}(x)}{1+x},n+1\right)=-(n+1)\mathrm{M}(\mathrm{H}_{-1,\boldsymbol{m}}(x),n)+\mathrm{H}_{-1,\boldsymbol{m}}(1),
\end{equation}
\begin{equation}
    \begin{aligned}
\mathrm{M}\left(\mathrm{H}_0(x), n\right) & =-\frac{1}{(n+1)^2} \\
\mathrm{M}\left(\mathrm{H}_1(x), n\right) & =\frac{\mathrm{S}_1(n+1)}{n+1} \\
\mathrm{M}\left(\mathrm{H}_{-1}(x), n\right) & =(-1)^n \frac{\mathrm{S}_{-1}(n+1)}{n+1}+\frac{\mathrm{H}_{-1}(1)}{n+1}\left(1+(-1)^n\right)
\end{aligned}
\end{equation}

\begin{equation}
    \begin{aligned}
\mathrm{M}\left(\mathrm{H}_{0, \boldsymbol{m}}(x), n\right) & =\frac{\mathrm{H}_{0, \boldsymbol{m}}(1)}{n+1}-\frac{1}{n+1} \mathrm{M}\left(\mathrm{H}_{\boldsymbol{m}}(x), n\right) \\
\mathrm{M}\left(\mathrm{H}_{1, \boldsymbol{m}}(x), n\right) & =\frac{1}{n+1} \sum_{i=0}^n \mathrm{M}\left(\mathrm{H}_m(x), n\right) \\
\mathrm{M}\left(\mathrm{H}_{-1, \boldsymbol{m}}(x), n\right) & =\frac{1+(-1)^n}{n+1} \mathrm{H}_{-1, \boldsymbol{m}}(1)-\frac{(-1)^n}{n+1} \sum_{i=0}^n(-1)^i \mathrm{M}\left(\mathrm{H}_m(x), n\right).
 \end{aligned}
\end{equation}
According to these properties, $\mathrm{M}(\mathrm{H}_{\boldsymbol{m}}(x),n)$, $\mathrm{M}(\frac{\mathrm{H}_{\boldsymbol{m}}(x)}{1-x},n)$ and $\mathrm{M}(\frac{\mathrm{H}_{\boldsymbol{m}}(x)}{1+x},n)$ can be recursively calculated and expressed in terms of the harmonic sums with the argument $n$, the harmonic sums at infinity and harmonic polylogarithms at one. Inversely, for specific harmonic sum $\mathrm{S}_{\boldsymbol{a}}(n)$, we can find $\frac{\mathrm{H}_{\boldsymbol{m}}(x)}{1+\omega x}$ such that $\mathrm{S}_{\boldsymbol{a}}(n)$ is the most complicated term in $\mathrm{M}(\frac{\mathrm{H}_{\boldsymbol{m}}(x)}{1+\omega x},n)$, i.e.,
\begin{equation}
    \mathrm{S}_{\boldsymbol{a}}(n)=\mathrm{M}\left(\frac{\mathrm{H}_{\boldsymbol{m}}(x)}{1+\omega x},n\right)+\mathrm{T},
\end{equation} 
where $\mathrm{T}$ is an expression in less complicated harmonic sums and constants.
The detailed procedure is given by $Algoritm\ 2$ in \cite{Ablinger:2009ovq}. According to the relation above, the asymptotic behavior of $\mathrm{S}_{\boldsymbol{a}}(n)$ can be computed by the asymptotic expansion of $\mathrm{M}(\frac{\mathrm{H}_{\boldsymbol{m}}(x)}{1+\omega x},n)$. Here is an example:
\begin{equation}
    \mathrm{S}_{3,1}(n)=-\mathrm{M}\left(\frac{\mathrm{H}_{0,0,1}(x)}{1- x},n\right)-\mathrm{S}_1(n)\mathrm{S}_3(\infty)+\mathrm{S}_2(n)\mathrm{S}_2(\infty),
\end{equation}
\begin{equation}
    \mathrm{M}\left(\frac{\mathrm{H}_{0,0,1}(x)}{1-x},n\right)
    =\frac{\mathrm{S}_2(\infty)^2}{2}-\mathrm{S}_3(\infty)\mathrm{S}_1(n)-\frac{S_2(\infty)}{n}+\mathcal{O}(\frac{1}{n^2}).
\end{equation}
Thus,
\begin{equation}
    \mathrm{S}_{3,1}(n)=\frac{\mathrm{S}_2(\infty)^2}{2}+\mathcal{O}(\frac{1}{n^2}).
\end{equation}
Here we replace the value of harmonic polylogarithms at one with the value of harmonic sums at infinity, and reduce the constant value through relations between harmonic sums at infinity. These relations contain quasi shuffle relations and the following relation\cite{Vermaseren:1998uu}:
\begin{equation}
    \mathrm{S}_{m_1, \ldots, m_p}(\infty) \mathrm{S}_{k_1, \ldots, k_q}(\infty)=\lim _{n \rightarrow \infty} \sum_{i=1}^n \frac{\operatorname{sign}\left(k_1\right)^i \mathrm{~S}_{m_1, \ldots, m_p}(n-i) \mathrm{S}_{k_2, \ldots, k_q}(i)}{i^{\left|k_1\right|}}.
\end{equation}
Considering these relations, the value of harmonic sum at infinity can be reduced to basis constants. 

Some constant values of harmonic sums at infinity can be found in \cite{Alday:2013cwa}. 

\bibliographystyle{JHEP}
\bibliography{refs}

\end{document}